\documentclass[article,a4paper,oneside,10pt]{memoir}

\usepackage{amsmath}
\usepackage{amssymb}
\usepackage{graphicx}

\usepackage{amsthm}
\usepackage{amstext}
\usepackage{stmaryrd}

\usepackage{color}

\usepackage{pict2e}

\usepackage[pdftex,plainpages=false,colorlinks,hyperindex,bookmarksopen,linkcolor=gray,citecolor=blue,urlcolor=blue]{hyperref}

\usepackage{tikz}
\usetikzlibrary{shapes}
\usetikzlibrary{positioning}

\usepackage{subcaption}

\makeatletter
\DeclareRobustCommand{\lintprod}{%
  \mathbin{\mathpalette\int@prod{(0,0)(0.8,0)(0.8,0.6)}}%
}
\DeclareRobustCommand{\rintprod}{%
  \mathbin{\mathpalette\int@prod{(0.1,0.6)(0.1,0)(0.9,0)}}}

\newcommand{\int@prod}[2]{%
  \begingroup
  \sbox\z@{$\m@th#1+$}%
  \setlength\unitlength{\wd\z@}%
  \linethickness{0.09\unitlength}%
  \begin{picture}(1,1)
  \roundcap
  \polyline#2
  \end{picture}%
  \endgroup
}
\makeatother

\counterwithout{section}{chapter}
\settrimmedsize{297mm}{210mm}{*}
\settypeblocksize{664pt}{498.13pt}{*}
\setulmargins{3cm}{*}{*}
\setlrmargins{*}{*}{0.75}
\setmarginnotes{17pt}{51pt}{\onelineskip}
\setheadfoot{\onelineskip}{2\onelineskip}
\setheaderspaces{*}{2\onelineskip}{*}
\checkandfixthelayout

\usepackage[justification=centerlast]{caption}
\usepackage{bm}

\newcommand{\Ab}{{\mathbf A}}
\newcommand{\Bb}{{\mathbf B}}

\newcommand{\Eb}{{\mathbf E}}
\newcommand{\Fb}{{\mathbf F}}

\newcommand{\Jb}{{\mathbf J}}

\newcommand{\Rb}{{\mathbf R}}
\newcommand{\Sb}{{\mathbf S}}
\newcommand{\Tb}{{\mathbf T}}
\newcommand{\ebf}{{\mathbf e}}

\newcommand{\fb}{{\mathbf f}}
\newcommand{\fbit}{\pmb{\mathit f}}

\newcommand{\jb}{{\mathbf j}}

\newcommand{\xb}{{\mathbf x}}

\newcommand{\ub}{{\mathbf u}}
\newcommand{\ubf}{{\mathbf u}}
\newcommand{\vb}{{\mathbf v}}
\newcommand{\wb}{{\mathbf w}}
\newcommand{\drm}{\,\mathrm{d}}

\newcommand{\deltabf}{{\boldsymbol \partial}}
\newcommand{\xibf}{{\boldsymbol \xi}}

\newcommand{\zetabf}{{\boldsymbol \zeta}}

\newcommand{\mf}{{\mathbf F}}
\newcommand{\mfi}{F}

\newcommand{\mff}{{\mathbf {\hat{F}}}}
\newcommand{\sd}{{\mathbf J}}
\newcommand{\sdf}{{\mathbf {\hat{J}}}}
\newcommand{\vp}{{\mathbf A}}

\newcommand{\vpf}{\hat{\mathbf {A}}}

\newcommand{\gf}{{\mathbf G}}

\newcommand{\tens}{\mathbf T}
\newcommand{\tensc}{T}

\DeclareMathOperator{\Tr}{Tr}

\DeclareMathOperator{\sgn}{sgn}
\DeclareMathOperator{\gr}{gr}

\newcommand{\len}[1]{\lvert#1\rvert}

\newcommand{\hodge}{{\scriptscriptstyle\mathcal{H}}}
\newcommand{\hodgeinv}{{\scriptscriptstyle\mathcal{H}^{-1}}}

\makeatletter
\let\save@mathaccent\mathaccent
\newcommand*\if@single[3]{%
  \setbox0\hbox{${\mathaccent"0362{#1}}^H$}%
  \setbox2\hbox{${\mathaccent"0362{\kern0pt#1}}^H$}%
  \ifdim\ht0=\ht2 #3\else #2\fi
  }
\newcommand*\rel@kern[1]{\kern#1\dimexpr\macc@kerna}
\newcommand*\widebar[1]{\@ifnextchar^{{\wide@bar{#1}{0}}}{\wide@bar{#1}{1}}}
\newcommand*\wide@bar[2]{\if@single{#1}{\wide@bar@{#1}{#2}{1}}{\wide@bar@{#1}{#2}{2}}}
\newcommand*\wide@bar@[3]{%
  \begingroup
  \def\mathaccent##1##2{%
    \let\mathaccent\save@mathaccent
    \if#32 \let\macc@nucleus\first@char \fi
    \setbox\z@\hbox{$\macc@style{\macc@nucleus}_{}$}%
    \setbox\tw@\hbox{$\macc@style{\macc@nucleus}{}_{}$}%
    \dimen@\wd\tw@
    \advance\dimen@-\wd\z@
    \divide\dimen@ 3
    \@tempdima\wd\tw@
    \advance\@tempdima-\scriptspace
    \divide\@tempdima 10
    \advance\dimen@-\@tempdima
    \ifdim\dimen@>\z@ \dimen@0pt\fi
    \rel@kern{0.6}\kern-\dimen@
    \if#31
      \overline{\rel@kern{-0.6}\kern\dimen@\macc@nucleus\rel@kern{0.4}\kern\dimen@}%
      \advance\dimen@0.4\dimexpr\macc@kerna
      \let\final@kern#2%
      \ifdim\dimen@<\z@ \let\final@kern1\fi
      \if\final@kern1 \kern-\dimen@\fi
    \else
      \overline{\rel@kern{-0.6}\kern\dimen@#1}%
    \fi
  }%
  \macc@depth\@ne
  \let\math@bgroup\@empty \let\math@egroup\macc@set@skewchar
  \mathsurround\z@ \frozen@everymath{\mathgroup\macc@group\relax}%
  \macc@set@skewchar\relax
  \let\mathaccentV\macc@nested@a
  \if#31
    \macc@nested@a\relax111{#1}%
  \else
    \def\gobble@till@marker##1\endmarker{}%
    \futurelet\first@char\gobble@till@marker#1\endmarker
    \ifcat\noexpand\first@char A\else
      \def\first@char{}%
    \fi
    \macc@nested@a\relax111{\first@char}%
  \fi
  \endgroup
}
\makeatother

\makeatletter
    \def\@endtheorem{\hfill$\P$\endtrivlist\@endpefalse }
\makeatother

\begin{document}

\title{Generalized Maxwell equations for exterior-algebra multivectors in $(k,n)$ space-time dimensions
\thanks{This work has been funded in part by the Spanish Ministry of Science, Innovation and Universities under grants TEC2016-78434-C3-1-R and BES-2017-081360. Published in: \textbf{\textit{Eur. Phys. J. Plus} 135, 305 (2020). DOI:} \href{https://doi.org/10.1140/epjp/s13360-020-00305-y}{10.1140/epjp/s13360-020-00305-y}.}
}

\author{\scshape Ivano Colombaro\thanks{ivano.colombaro@upf.edu}, Josep Font-Segura\thanks{josep.font@ieee.org}, Alfonso Martinez\thanks{alfonso.martinez@ieee.org}\thanks{Authors are with the Department of Information and Communication Technologies, Universitat Pompeu Fabra, Barcelona, Spain.}}

\maketitle

\begin{abstract}
This paper presents an exterior-algebra generalization of electromagnetic fields and source currents as multivectors of grades $r$ and $r-1$ respectively in a flat space-time with $n$ space and $k$ time dimensions. Formulas for the Maxwell equations and the Lorentz force for arbitrary values of $r$, $n$, and $k$ are postulated in terms of interior and exterior derivatives, in a form that closely resembles their vector-calculus analogues. These formulas lead to solutions in terms of potentials of grade $r-1$, and to conservation laws in terms of a stress-energy-momentum tensor of rank 2 for any values of $r$, $n$, and $k$, for which a simple explicit formula is given. As an application, an expression for the flux of the stress-energy-momentum tensor across an $(n+k-1)$-dimensional slice of space-time is given in terms of the Fourier transform of the potentials.
The abstraction of Maxwell equations with exterior calculus combines the simplicity and intuitiveness of vector calculus, as the formulas admit explicit expressions, with the power of tensors and differential forms, as the formulas can be given for any values of $r$, $n$, and $k$.
\end{abstract}



\section{Introduction}

Classical electromagnetic phenomena are described by solutions of Maxwell's equations, a set of coupled partial differential equations relating the electric and magnetic fields across space and time with the charges and currents generating them. Let $\Eb(t,\xb)$ and $\Bb(t,\xb)$ respectively denote the electric and magnetic vector fields distributed over time $t$ and space $\xb$, and $\rho(t,\xb)$ and $\jb(t,\xb)$ respectively denote the charge and current densities. In vector-calculus notation, and in units chosen such that the speed of light is set to 1 and all the quantities in the pairs time and space, electric and magnetic fields, and charge and current density have the same units, the standard microscopic Maxwell's equations take the well-known form \cite[Sects.~26,~30]{landau1982classicalTheoryFields} or \cite[Sect.~6.6]{jackson1999classicalElectrodynamics}
\begin{gather}
\nabla\cdot\Eb = \rho \label{eq:maxEq_intro_1} \\
\nabla\times\Eb = -\frac{\partial\Bb}{\partial t} \label{eq:maxEq_intro_2} \\
\nabla\cdot\Bb = 0 \label{eq:maxEq_intro_3} \\
\nabla\times\Bb = \jb + \frac{\partial\Eb}{\partial t}. \label{eq:maxEq_intro_4}
\end{gather}
Conventionally, Eqs.~\eqref{eq:maxEq_intro_2} and~\eqref{eq:maxEq_intro_3} (resp.~\eqref{eq:maxEq_intro_1} and~\eqref{eq:maxEq_intro_4}) are referred to as homogeneous (resp.~inhomogeneous) Maxwell equations.

As a complement to Maxwell equations, the electromagnetic field creates in turn a force on charge and current densities described by the Lorentz force density $\fbit$ \cite[Sect.~17]{landau1982classicalTheoryFields} or \cite[Sect.~6.7]{jackson1999classicalElectrodynamics}, a vector field given by
\begin{equation}
\fbit = \rho \Eb + \jb\times\Bb.
\end{equation}
Integrated over a region of space, the Lorentz force density $\fb(t,\xb)$ gives the force acting upon the charges in that region. In addition, the rate of work being done on the charges by the fields is given by  $\jb\cdot \Eb$ \cite[Sect.~17]{landau1982classicalTheoryFields} or \cite[Sect.~11.9]{jackson1999classicalElectrodynamics}. 

In relativistic form, the charge and current densities are combined into a single vector $\Jb$ with four components whose time and space components are $\rho$ and $\jb$ respectively, while the electric and magnetic fields are combined into an antisymmetric tensor $\Fb$ of rank 2 indexed by pairs of space-time dimensions, the Faraday tensor. In the space-time metric $(-1,1,1,1)$, the components $\alpha,\beta$ of the Faraday tensor in its contravariant form, $F^{\alpha\beta}$ \cite[Sect.~23]{landau1982classicalTheoryFields} or \cite[Sect.~6.7]{jackson1999classicalElectrodynamics}, are given by
\begin{equation}\label{eq:faraday-tensor}
	F^{\alpha\beta} = \begin{pmatrix} 0 & E_x & E_y & E_z \\ -E_x & 0 & B_z & -B_y \\ -E_y & -B_z & 0 & B_x \\  -E_z & B_y & -B_x & 0 \end{pmatrix}.
\end{equation}
Both pairs of homogeneous and inhomogeneous equations are combined into single equations, namely \cite[Sects.~26, 30]{landau1982classicalTheoryFields} or \cite[Sect.~11.9]{jackson1999classicalElectrodynamics}
\begin{gather}
	\partial_\alpha F_{\beta\gamma} + \partial_\beta F_{\gamma\alpha} + \partial_\gamma F_{\alpha\beta} = 0 \\
	\partial_\alpha F^{\alpha\beta} = J^\beta.
\end{gather}
Also, the Lorentz force becomes a four-dimensional vector $\fb$, whose time and space components are $\jb\cdot \Eb$ and $\fbit$ respectively. The component $\alpha$ of the Lorentz force, $f^{\alpha}$, is given in its contravariant form by \cite[Sect.~29]{landau1982classicalTheoryFields} or \cite[Sect.~11.9]{jackson1999classicalElectrodynamics}
\begin{equation}\label{eq:lorentz-force-tens}
	\text{f}^{\alpha} = F^{\alpha\beta}J_\beta.
\end{equation} 

The basic equations of electromagnetism have a very rich history beyond the common vector and tensor formulas given above. At various moments quaternions \cite{maxwell1873treatiseElectricityMagnetism}, six-vectors \cite{sommerfeld1910zurRelativitaetstheorieI,sommerfeld1910zurRelativitaetstheorieII,wilson1911spacetimeManifoldRelativity}, geometric or Clifford algebras \cite{kaehler1937bermerkungenMaxwellschenGleinchungen,hestenes2015spacetimeAlgebra}, or differential forms \cite{edelen1978metricFreeElectrodynamics,deschamps1981electromagnetismDifferentialForms} have all been put forward as convenient descriptions of fields, sources, and forces. Depending on whether simplicity, generality, computation ease, or intuitiveness or some combination thereof is preferred, each of these objects carries its own advantages and disadvantages. The purpose of this paper is to show how a formulation by means of exterior algebra, building on historical work on six-vectors for the electromagnetic field by Sommerfeld \cite{sommerfeld1910zurRelativitaetstheorieI,sommerfeld1910zurRelativitaetstheorieII}, attains a good balance of simplicity in terms of notation and operations involved, with no need of covariant and contravariant representations; generality, as the equations are naturally cast in a relativistic form valid for any flat space-time with generalized electromagnetic fields and source densities given by $r$-vectors and $(r-1)$-vectors respectively, objects whose appearance is obscured when differential forms are used; and computation ease and intuitiveness shared with three-dimensional vector calculus.

In Sect.~\ref{sec:exterior} we present the necessary notions and operations of exterior algebra and calculus, as needed for our purposes. Explicit formulas for the exterior and interior products and derivatives of multivectors, and circulation and flux of multivector fields are given. These concepts are put into use in Sect.~\ref{sec:maxwell}, where we present the generalized Maxwell equations and Sect.~\ref{sec:lorentz}, presenting the generalized Lorentz force and a simple expression of its associated stress-energy tensor. Both sections include several applications to support 
exterior algebra as a natural setting for an abstract yet intuitively simple form of Maxwell equations, Lorentz force, and stress-energy tensor for any values of the grade $r$ and the space-time dimensions.

\section{Exterior Algebra and Calculus}
\label{sec:exterior}

In this section, we present the basic notions and operations of exterior calculus; the section contents are a condensed summary of \cite{colombaro2019introductionSpaceTimeExteriorCalculus}.

\subsection{Exterior Algebra: Space-time, Multivectors, and Products}

We build our theory on a flat space-time with $k$ time dimensions and $n$ space dimensions identified with $\Rb^{k+n}$. We represent the canonical basis of this $(k,n)$- or $(k+n)$-space-time  by $\smash{\{\ebf_i\}_{i=0}^{k+n-1}}$ and adopt the convention that its first $k$ indices, i.~e.~$i=0,\dotsc,k-1$, correspond to time components while the indices $i = k,\dotsc,k+n-1$ represent space components. A point or position in space-time is denoted by $\xb$, with components $x_i$ in the canonical basis $\smash{\{\ebf_i\}_{i=0}^{k+n-1}}$, that is 
\begin{equation}
\xb = \sum_{i=0}^{k+n-1}x_i\ebf_i	.
\end{equation}

As first observed by Grassmann \cite{grassmann1862extensionTheory}, next to the $(k+n)$-dimensional Euclidean vector space $\Rb^{k+n}$, for every value of the parameter $m = 0, \dotsc, k+n$ there exist other natural vector spaces with basis vectors $\ebf_I$ indexed by ordered lists $I = (i_1,\dotsc,i_m)$ of $m$ non-identical space-time indices. 
For instance, if $m = 0$, the list is empty and the vector space is $\Rb$; for $m = 1$ we recover the standard vector space $\Rb^{k+n}$; for $m = 2$ and $m = 3$, the exterior-algebra basis vectors can be respectively identified with oriented plane and volume elements. This parameter $m$ plays a very important role in exterior algebra and we refer to it as grade and to these vectors as multivectors or grade-$m$ vectors if we wish to be more specific. 
A general multivector field $\vb(\xb)$ of grade $m$, possibly a function of the position $\xb$, with components $v_I(\xb)$ in the canonical basis $\smash{\{\ebf_I\}_{I\in\mathcal{I}_m}}$ can be written as 
\begin{equation}\label{eq:multivector}
\vb(\xb) = \sum_{I\in\mathcal{I}_m} v_I(\xb)\ebf_I,
\end{equation}
where $\mathcal{I}_m$ denotes the set of all ordered lists with $m$ indices. The dimension of the vector space containing all grade-$m$ multivectors is $\smash{\binom{k+n}{m}}$, the number of lists in $\mathcal{I}_m$. We denote by $\len{I}$ the length of a list $I$, e.~g.~$\len{I} = m$ for $I \in \mathcal{I}_m$ and by $\gr(\vb)$ the operation that returns the grade of the vector $\vb$, e.~g.~$\gr{\vb} = m$ in~\eqref{eq:multivector}.

As Sommerfeld observed \cite{sommerfeld1910zurRelativitaetstheorieI,sommerfeld1910zurRelativitaetstheorieII}, the electric and magnetic fields may be combined in a six-vector with components indexed by pairs of distinct space-time dimensions. Rather than the usual antisymmetric Faraday tensor in~\eqref{eq:faraday-tensor}, one has a bivector, i.~e.~a multivector with grade 2. We extend this observation and consider a generalized electromagnetic field $\mf(\xb)$ at every space-time point $\xb$ given by an $r$-vector, of grade $r$, and a source density given by a vector $\sd (\xb)$ of grade $(r-1)$. We give this generalized electromagnetic field the name of Maxwell field. The phenomenological description of the charges corresponding to this source density $\sd(\xb)$ will be done elsewhere.
As for the generalized Lorentz force density $\fb(\xb)$, 
we shall see in Sect.~\ref{sec:lorentz} that it remains a vector of grade 1 with $k+n$ components. For convenience, we often write $\mf$, $\sd$, and $\fb$, dropping the explicit dependence on $\xb$. 

Having defined multivectors and their canonical bases for various grades $m$, we consider the operations acting on them. While these operations may change the grades of their input objects or have different inputs with different grades, we always consider objects with a fixed grade. In other words, these objects are not sums of multivectors with different grades, as it would be in a Clifford or geometric algebra \cite{hestenes1982cliffordAlgebra}.  
In order of presentation, and loosely following \cite[Sect.~2]{colombaro2019introductionSpaceTimeExteriorCalculus}, we define the dot or scalar product, the wedge product, the interior products, and the Hodge and inverse Hodge complements. With no real loss of generality, we define the operations only for the canonical basis vectors, the operation acting on general multivectors being a mere extension by linearity of the former.

Given two arbitrary canonical basis vectors $\ebf_{i}$ and $\ebf_{j}$ in $\Rb^{k+n}$, their dot space-time product, also denoted by $\Delta_{ij} = \ebf_{i}\cdot\ebf_{j}$, is
\begin{equation}\label{eq:dot_vec_basis}
\ebf_{i}\cdot\ebf_{j} = \begin{cases} -1, & i = j, \, 0\leq i \leq k-1, \\ 
1, & i = j, \, k\leq i \leq k+n-1. \\
 0, & i \neq j, \end{cases}
\end{equation}
We adopt the convention that time unit vectors $\ebf_{i}$ have negative norm $\Delta_{ii} = -1$ and space unit vectors $\ebf_{i}$ have positive norm $\Delta_{ii} = +1$. We extend the definition of dot product $\cdot$ to arbitrary grade-$m$ basis vectors $\ebf_I$ and $\ebf_J$ as
\begin{equation}\label{eq:dot_multi}
	\ebf_I\cdot\ebf_J = \Delta_{IJ} = \Delta_{i_1 j_1}\Delta_{i_2 j_2}\dotsm\Delta_{i_m j_m},
\end{equation}	
where $I$ and $J$ are the ordered lists $I = (i_1,i_2,\dotsc,i_m)$ and $J = (j_1,j_2,\dotsc,j_m)$. 

The exterior algebra is defined on the direct sum of these vector spaces over all values of $m$, resulting in a larger vector space of dimension $2^{k+n}$. In this larger space, one could define a geometric product between two multivectors, possibly of mixed grade, and thus obtain its associated geometric or Clifford algebra \cite{hestenes1982cliffordAlgebra}. Using instead the tensor product of two vectors, one would obtain the tensor algebra, for which multivectors correspond to antisymmetric tensors of rank $m$. 
For the purposes of electromagnetism, there is no need of considering the full machinery of tensor or geometric algebras.  In the exterior algebra, the product between two multivectors is the exterior or outer product, which we define next. 

Let two basis vectors $\ebf_I$ and $\ebf_J$ have grades $m = \len{I}$ and $m' = \len{J}$. Let $(I,J) = \{i_1,\dotsc,i_m,j_1,\dotsc,j_{m'}\}$ be the concatenation of $I$ and $J$, let $\sigma(I,J)$ denote the signature of the permutation sorting the elements of this concatenated list of $\len{I}+\len{J}$ indices, and let $I+J$, or $\varepsilon(I,J)$ if $I+J$ is ambiguous, denote the resulting sorted list. Then, the exterior product $\ebf_I$ of $\ebf_J$ is defined as
\begin{equation} \label{eq:ext-prod-def}
	\ebf_I\wedge\ebf_J = \sigma(I,J)\ebf_{I+J}.
\end{equation}
Since permutations with repeated indices have zero signature, the exterior product is zero if \mbox{$\len{I} + \len{J} > k+n$} or more generally if both vectors have at least one index in common. The exterior product is thus either zero or a vector of grade $\len{I}+\len{J}$. 

The exterior product constructs the basis vectors for any grade $m$ from the canonical basis vectors $\ebf_i$. We do so by identifying the vector $\ebf_I$ for the ordered list $I = (i_1,\dotsc,i_m)$ with the exterior product of $\ebf_{i_1}, \ebf_{i_2} ,\dotsc, \ebf_{i_m}$, that is 
\begin{equation}
	\ebf_I = \ebf_{i_1}\wedge\ebf_{i_2}\wedge\dotsm\wedge\ebf_{i_m}.
\end{equation}

Furthermore, as we can intuitively note from~\eqref{eq:ext-prod-def}, the exterior product is an operation that adds the grades of the input multivectors, while the dot product subtracts their grades, yielding a scalar, i.~e.~a zero-grade multivector. 
We now define two generalizations of the dot product, the left and right interior products of two multivectors, as grade-lowering operations that output a multivector whose grade is the difference of the input multivector grades.

Let $\ebf_I$ and $\ebf_J$ be two basis vectors of respective grades $\len{I}$ and $\len{J}$. The left interior product, denoted by $\lintprod$, is defined as
\begin{equation}
	\ebf_I \lintprod \ebf_J = \Delta_{II}\sigma(J\setminus I,I)\ebf_{J\setminus I}
	\label{eq:left-int-prod}
\end{equation}
if $I$ is a subset of $J$ and zero otherwise. The new basis $\ebf_{J\setminus I}$ is a vector of grade $\smash{\len{J}-\len{I}}$ and it presents the indices of $J$ excluding those in common with $I$.
The right interior product, denoted by $\rintprod$, of two basis vectors $\ebf_I$ and $\ebf_J$ is defined as 
\begin{equation}
	\ebf_I \rintprod \ebf_J = \Delta_{JJ}\sigma(J,I\setminus J)\ebf_{I\setminus J} \label{eq:right-int-prod}
\end{equation}
if $J$ is a subset of $I$ and zero otherwise.
Both are grade-lowering operations, as the left (resp.~right) interior product is either zero or a multivector of grade $\smash{\len{J}-\len{I}}$ (resp.~$\len{I}-\len{J}$).


Finally, we define the complement of a multivector, a concept needed to capture the idea of orthogonal subspace. For a unit vector $\ebf_I$ with grade $\len{I}$, its Grassmann or Hodge complement, denoted by $\ebf_I^\hodge$, is a unit $(k+n-\len{I})$-vector given by
\begin{equation} \label{eq:hodge-transf}
	\ebf_I^\hodge = \Delta_{II}\sigma(I,I^c)\ebf_{I^c},
\end{equation}
where $\smash{I^c}$ is the complement of the list $I$, namely the ordered sequence of indices not included in $I$, and $\sigma(I,I^c)$ is the signature of the permutation sorting the elements of the concatenated list $(I,I^c)$ of all space-time indices.
In other words $\smash{\ebf_{I^c}}$ is a basis vector of grade $k+n-\len{I}$ whose indices are absent from $I$. In addition, we define the inverse complement transformation as
\begin{equation} \label{eq:hodge-inv-transf}
	\ebf_I^\hodgeinv = \Delta_{I^c I^c}\sigma(I^c,I)\ebf_{I^c} .
\end{equation}


\subsection{Some Properties of the Exterior and Interior Products}

This section lists some relevant commutative and distributive properties of the exterior and interior products of multivectors of fixed grade; these properties will be needed later in our generalized electromagnetic sources, fields, and equations. 

The exterior product~\eqref{eq:ext-prod-def} is a skew-commutative operation, as we have
\begin{equation}\label{eq:comm-extprod}
	\ub\wedge\vb = (-1)^{\gr{\ub}\cdot\gr{\vb}}\vb\wedge\ub.
\end{equation}

Concerning the interior products, we have that
\begin{equation}\label{eq:comm-intprod}
\ub \lintprod \vb = \vb \rintprod \ub (-1)^{\gr{\ub}\cdot (\gr{\ub}+\gr{\vb})},
\end{equation}
and the interior products are also skew-commutative, unless $\gr{\ub} = \gr{\vb}$, when both are commutative and coincide with the dot product of the two vectors.

Given two vectors $\vb$ and $\vb'$ and two $r$-vectors $\wb$ and $\wb'$, then it holds that 
\begin{equation} \label{eq:text-equiv-wedge-int}
\left( \vb \wedge \wb \right) \cdot \left( \wb' \wedge \vb' \right) = (-1)^r \left( \vb \cdot \vb' \right) \left( \wb \cdot \wb' \right)
+ \left( \vb' \lintprod \wb \right) \cdot \left( \wb' \rintprod \vb \right).
\end{equation}
Or equivalently, using~\eqref{eq:comm-extprod} and~\eqref{eq:comm-intprod}, it holds that
\begin{equation}
\left( \vb \cdot \vb' \right) \left( \wb \cdot \wb' \right) =
\left( \vb \wedge \wb \right) \cdot \left( \vb' \wedge \wb' \right) +
\left( \vb \lintprod \wb' \right) \cdot \left( \vb' \lintprod \wb \right) .
\end{equation}

Consider now two vectors $\ub$ and $\vb$ and a $r$-vector $\wb$, then it holds that 
\begin{equation} \label{eq:text-equiv-double-interior}
\ub \lintprod (\wb \rintprod \vb) = (\ub \lintprod \wb) \rintprod \vb ,
\end{equation}
or alternatively, using~\eqref{eq:comm-intprod}, we have that
\begin{equation} \label{eq:text-equiv-double-interior2}
\ub \lintprod (\vb \lintprod \wb)  = -  \vb \lintprod (\ub \lintprod \wb).
\end{equation}
As already found in~\cite[Eq.~(16)]{colombaro2019introductionSpaceTimeExteriorCalculus}, for the same  $\ub$, $\vb$, and $\wb$ it also holds that
\begin{equation} \label{eq:equiv-lint-wedge}
\ub \lintprod (\vb \wedge \wb) = (-1)^r (\ub\cdot\vb) \wb + \vb \wedge (\ub \lintprod \wb) .
\end{equation}

Finally, for a vector $\ub$, a $(r-1)$-vector $\vb$ and a $r$-vector $\wb$, the following equalities are true 
\begin{equation} \label{eq:text-wedge-dot-}
(\ub \wedge \vb) \cdot \wb = \vb \cdot (\wb \rintprod \ub) = \ub \cdot (\vb \lintprod \wb),
\end{equation}
or, alternatively, using~\eqref{eq:comm-intprod}, we have that
\begin{equation} 
(\ub \wedge \vb) \cdot \wb = (-1)^{r-1} (\ub \lintprod \wb) \cdot \vb  = (\vb \lintprod \wb) \cdot \ub.
\end{equation}

The relations~\eqref{eq:text-equiv-wedge-int},~\eqref{eq:text-equiv-double-interior} and~\eqref{eq:text-wedge-dot-}
are proved in~\ref{app:properties}.

\subsection{Exterior Calculus: Derivatives and Integrals, Circulation and Flux} 

In vector calculus, extensive use is made of the nabla operator $\nabla$, a vector operator that takes partial space derivatives. For example, the divergence and curl in Maxwell equations are expressed in terms of this operator. In our case, we need its generalization to $(k,n)$-space-time,  the differential vector operator $\deltabf$ defined as  $(-\partial_0,-\partial_2,\dotsc,-\partial_{k-1}, \partial_{k},\dotsc,\partial_{k+n-1})$, that is
\begin{equation}
	\deltabf = \sum_{i=0}^{k+n-1}\Delta_{ii}\ebf_i\partial_i.
\end{equation}	

As done in more detail in \cite[Sect.~3]{colombaro2019introductionSpaceTimeExteriorCalculus}, we define the exterior derivative of $\vb$, $\deltabf\wedge \vb$, of a given vector field $\vb$ of grade $m$ as
\begin{equation}\label{eq:ext-deriv-def}
\deltabf\wedge \vb = \sum_{i=0}^{k+n-1} \sum_{I\in\mathcal{I}_m} \Delta_{ii}\sigma(i,I)\partial_i v_I  \, \ebf_{i+I} = \sum_{i,I\in\mathcal{I}_m:\, i\notin I} \Delta_{ii}\sigma(i,I)\partial_i v_I  \, \ebf_{i+I}.
\end{equation}
The grade of the exterior derivative of $\vb$ is $m+1$, unless $m = k+n$, in which case the exterior derivative is zero, as can be deduced from the fact that all signatures are zero.
In addition, we define the interior derivative of $\vb$ as $\deltabf\lintprod \vb$, namely
\begin{equation}\label{eq:int-deriv-def}
\deltabf\lintprod \vb = \sum_{i=0}^{k+n-1} \sum_{I\in\mathcal{I}_m}\sigma(I\setminus i,i)  \partial_i v_I \ebf_{I\setminus i}  = 
\sum_{i,I\in\mathcal{I}_m:\, i\in I} \sigma(I\setminus i,i) \partial_i v_I \ebf_{I\setminus i} .
\end{equation}
The grade of the interior derivative of $\vb$ is $m-1$, unless $m = 0$, in which case the interior derivative is zero, as can be deduced from the fact the grade of $\deltabf$ is larger than the grade of $\vb$.
It is easy to verify that the exterior derivative of an exterior derivative is zero, as is the interior derivative of an interior derivative, that is for any vector field $\vb$, we have that 
\begin{align}
	\deltabf \wedge (\deltabf \wedge \vb) &= 0 \label{eq:ext-der-ext-der} \\
	\deltabf \lintprod (\deltabf \lintprod \vb) &= 0. \label{eq:int-der-int-der}
\end{align}
From the identity~\eqref{eq:equiv-lint-wedge}, we note that the interior derivative of the exterior derivative is related to the exterior derivative of the interior derivative as
\begin{equation}\label{eq:left-wedge}
	\deltabf\lintprod(\deltabf\wedge\vb ) = (-1)^{\gr(\vb)}(\deltabf\cdot\deltabf)\vb + \deltabf\wedge(\deltabf\lintprod\vb).
\end{equation}

From the chain of identities in~\eqref{eq:text-wedge-dot-}, given a $(r-1)$-vector $\vb$ and a $r$-vector $\wb$, then the equalities convert into the product rule for the derivative which we write
\begin{equation}
\deltabf \cdot (\vb \lintprod \wb) = (\deltabf \wedge \vb) \cdot \wb + (-1)^{\gr{(\vb)}} (\deltabf \lintprod \wb)\cdot \vb ,
\end{equation} 
taking into account that the operator $\deltabf$ is described by a vector basis $\ebf_i$ but it acts as a derivative, too.

The formulas for the exterior and interior derivatives allow us express some common expressions in vector calculus. For instance, the gradient of a scalar field $\omega$ is given by $\nabla\omega = \deltabf\wedge\omega$ or the divergence of a vector field $\vb$  is given by $\nabla\cdot\vb = \deltabf\lintprod \vb$. As another example, the cross product of two vector fields $\vb$ and $\wb$ in $\Rb^3$ can be variously expressed as
\begin{equation}\label{eq:34}
	\vb\times\wb = (\vb\wedge\wb)^\hodgeinv = \vb\lintprod\wb^\hodgeinv = \vb\lintprod\wb^\hodge.
\end{equation}
This formula allows us to write the curl of a three-dimensional vector field in several equivalent ways 
in terms of the exterior and interior products and the Hodge complements, thereby providing a generalization of the cross product and the curl to grade-$m$ vector fields in space-times with different dimensions. 

Integrals are, together with derivatives, the fundamental mathematical objects of calculus. For example, operations on vectors fields such as flux and circulation are expressed in terms of integrals over high-dimensional geometric objects. For any $\ell = 0,\dotsc,k+n$, we 
define an infinitesimal vector element $\drm^{\ell}\xb$ as the sum of all possible differentials for $\ell$-dimensional hypersurfaces in a $(k,n)$-space-time. This infinitesimal vector element is represented in the canonical basis as
\begin{equation} \label{eq:differential-element}
	\drm^{\ell}\xb = \sum_{I\in\mathcal{I}_\ell}\drm x_{I}\ebf_I,
\end{equation}
where for a given list $I=(i_1,\dotsc,i_\ell)$ the differential is given by $dx_{I}=\drm x_{i_1}\cdots \drm x_{i_\ell}$. 
A positive orientation is implicit in~\eqref{eq:differential-element}, as the skew-symmetry of the product~\eqref{eq:ext-prod-def} may introduce sign changes to compensate an eventual change of orientation after coordinates change, e.~g.~permutations of the space-time components. 

The circulation of a vector field $\vb(\xb)$ of grade $m$ along an $\ell$-dimensional hypersurface $\mathcal{V}^\ell$ with $\ell = m$ is defined in terms of the right interior product, or equivalently the scalar product as $\ell = m$, that is
\begin{gather}
	\int_{\mathcal{V}^m}\drm^{m}\xb\rintprod\vb = \int_{\mathcal{V}^m}\drm^{m}\xb\cdot\vb.
	\label{eq:def1}
\end{gather}
This formula recovers for $\ell = m = 1$ and $\Rb^n$ the definition of the circulation of a vector field along a closed path with the appropriate orientation. 
Intuitively, the circulation measures the alignment of the vector field $\vb$ with respect to ${\mathcal V}^m$.

The Stokes theorem for the circulation \cite[Sect.~3.4]{colombaro2019introductionSpaceTimeExteriorCalculus} states that the circulation of a grade-$m$ vector field $\vb$ along the boundary $\partial\mathcal{V}^{m+1}$ of an $(m+1)$-dimensional hypersurface $\mathcal{V}^{m+1}$ is equal to the circulation of the exterior derivative of $\vb$ along the hypersurface $\mathcal{V}^{m+1}$:
\begin{equation} \label{eq:circulation}
	\int_{\partial\mathcal{V}^{m+1}}\drm^{m}\xb\cdot\vb = \int_{\mathcal{V}^{m+1}}\drm^{m+1}\xb\cdot(\deltabf\wedge \vb).
\end{equation}
The role of the vector curl in the usual form of the Kelvin-Stokes theorem is played by the exterior derivative in this version of the theorem. Recall that the Kelvin-Stokes theorem for the circulation of a vector field $\vb$ of grade 1 along the boundary $\partial\mathcal{V}^{2}$ of a bidimensional surface $\mathcal{V}^{2}$ relates its value to that of the surface integral of the curl of the vector field over the surface itself. 


The flux of a vector field $\vb(\xb)$ of grade $m$ across an $\ell$-dimensional hypersurface $\mathcal{V}^\ell$ is defined in terms of the left interior product and the inverse Hodge complement of the infinitesimal vector element~\eqref{eq:differential-element}, that is
\begin{gather}
	\int_{\mathcal{V}^\ell}\drm^{\ell}\xb^{\hodgeinv}\lintprod\vb.
	\label{eq:def2}
\end{gather}
For example, this formula recovers the flux across a line on the plane or across a surface in three-dimensional space. 
Alternatively, if $\ell = k + n$, one can verify that
the flux of $\vb$ over an $(k+n)$-dimensional hypersurface $\mathcal{V}^{k+n}$ gives the volume integral of $\vb$ over $\mathcal{V}^{k+n}$.
Intuitively, the flux~\eqref{eq:def2} measures the magnitude of the multivector field crossing the hypersurface. In general, the flux is a vector of grade $(m+\ell-n-k)$ if $\ell \geq k+n-m$ and zero otherwise. 

As with the circulation, a generalized Stokes theorem \cite[Sect.~3.5]{colombaro2019introductionSpaceTimeExteriorCalculus} states that the flux of a grade-$m$ vector field $\vb$ across the boundary $\partial\mathcal{V}^{\ell}$ of an $\ell$-dimensional hypersurface $\mathcal{V}^{\ell}$ is equal to the flux of the interior derivative of $\vb$ across $\mathcal{V}^{\ell}$:
\begin{equation}\label{eq:flux-2}
	\int_{\partial\mathcal{V}^{\ell}}\drm^{\ell-1}\xb^{\hodgeinv}\lintprod\vb = 
	\int_{\mathcal{V}^{\ell}}\drm^{\ell}\xb^{\hodgeinv}\lintprod(\deltabf\lintprod \vb).
\end{equation}

Armed with the relevant definitions of space-time algebra, the products defined on multivectors, and the derivatives and integrals in exterior calculus, we postulate in the next section the generalized Maxwell equations for arbitrary $r$, $k$, and $n$.

\section{Generalized Maxwell Equations}
\label{sec:maxwell}


\subsection{Differential Form of Maxwell Equations}

For a given $r$, we consider a Maxwell field $\mf(\xb)$ and a generalized source density $\sd(\xb)$ at every point $\xb$ of the flat $(k,n)$-space-time. The Maxwell field $\mf(\xb)$ is a multivector field of grade $r$ and the source density $\sd(\xb)$ a multivector field of grade $(r-1)$. For convenience, we often write $\mf$ and $\sd$, dropping the explicit dependence on $\xb$. Going directly to the heart of the matter, we postulate the generalized Maxwell equations for arbitrary $r$, $k$, and $n$ to be the following two coupled differential equations relating $\mf(\xb)$ and $\sd(\xb)$:
\begin{gather} 
\deltabf\lintprod \mf = \sd, \label{eq:maxwell-Gen1}\\
\deltabf\wedge \mf = 0. \label{eq:maxwell-Gen2}
\end{gather}
The wedge product~\eqref{eq:ext-prod-def} in the exterior derivative~\eqref{eq:ext-deriv-def} raises the grade of $\Fb$ and the zero in~\eqref{eq:maxwell-Gen2} is the null $(r+1)$-vector; similarly, as the left interior product~\eqref{eq:left-int-prod} in the interior derivative~\eqref{eq:int-deriv-def} lowers the grade of $\Fb$, both sides of~\eqref{eq:maxwell-Gen1} are $(r-1)$-vectors. A source density conservation law,  $\deltabf\lintprod\Jb = 0$, follows from~\eqref{eq:maxwell-Gen1} and~\eqref{eq:int-der-int-der}.

For given $r$, the Maxwell field and the source density have respectively $\binom{k+n}{r}$ and $\binom{k+n}{r-1}$ components at each point of space-time. For $r=2$, $k = 1$, and $n = 3$, the field is the usual electromagnetic field, in bivector form rather than in the tensor form~\eqref{eq:faraday-tensor}, whose three components with space-space indices $F_{ij}(\xb)$, $i,j = 1,2,3$, $i<j$, represent the magnetic field, and whose remaining three components with space-time indices $F_{ij}(\xb)$, $i = 0$, $j = 1,2,3$, represent the electric field. More precisely, we have $\mf = \ebf_0\wedge\Eb + \Bb^{\hodge}$, with the Hodge complement defined in~\eqref{eq:hodge-transf}. 
Similarly, the first component of the source density represents the density of charge $\rho(\xb)$ and the last three represent the space current density $\jb(\xb)$, namely $\sd = \rho\ebf_0 + \jb$. It can be verified rather easily that these bivector equations are equivalent to the differential form of the vector Maxwell equations in~\eqref{eq:maxEq_intro_1}--\eqref{eq:maxEq_intro_4} \cite[Sect.~4.2]{colombaro2019introductionSpaceTimeExteriorCalculus}.

Two other less obvious examples are $r=1$, $k = 0$, and $n = 3$, and $r=2$, $k = 0$, and $n = 3$. In the first case, the field $\mf$ is a vector field that we can identify with $\Eb$, the source density is a scalar, namely $\rho$ and there is no time, and we recover the equations of electrostatics. Indeed, as $\deltabf = \nabla$ and $\deltabf\lintprod = \nabla \cdot$, and using~\eqref{eq:34} to write $\deltabf\wedge = \nabla \times$ in three dimensions, we obtain
\begin{gather} 
\nabla\cdot \Eb = \rho,\\
\nabla \times \Eb = 0. 
\end{gather}
In the second case, the field $\mf$ is a bivector field that we can identify with the Hodge complement of $\Bb$, that is $\mf = \Bb^\hodge$, the source density is a vector, namely $\jb$, and we recover the equations of magnetostatics. Indeed, using~\eqref{eq:34} we obtain
\begin{gather} 
\nabla\times \Bb = \jb,\\
\nabla \cdot \Bb = 0. 
\end{gather}
Polar vectors, such as the electric field, are naturally represented by a vector. 
Although axial vectors, such as the magnetic fied, can be represented as vectors in three dimensions, it might be more natural to represent them as bivectors.

\subsection{Vector Potential}
\label{sec:vector_potential}
As in standard electromagnetism, one can introduce a potential field $\vp(\xb)$, now a multivector field of grade $r-1$ with $\binom{k+n}{r-1}$ components. 
According to Poincar\'{e}'s Lemma \cite[Sect.~36.G]{arnold1989mathematicalMethods}, stated in exterior-algebra notation, whenever the exterior derivative of an $r$-vector field $\Fb$ vanishes on a contractible domain, then $\Fb$ must be the exterior derivative of an $(r-1)$ vector. In a suitably contractible domain, this Lemma implies that the homogenous Maxwell equation~\eqref{eq:maxwell-Gen2} is equivalent to 
\begin{equation}	\label{eq:field-potential}
	\mf = \deltabf\wedge\vp.
\end{equation}
For the simple examples listed in the previous section, the potential is respectively the usual relativistic 4-potential, the scalar potential in electrostatics and the vector potential in magnetostatics. 
If we replace the potential $\vp$ by a new field $\vp' = \vp + \deltabf\wedge\gf$, where $\gf$ is an $(r-2)$-vector gauge field, the homogenous Maxwell equation~\eqref{eq:maxwell-Gen2} is unchanged thanks to~\eqref{eq:ext-der-ext-der}. There is therefore some unavoidable ambiguity on the value of the vector potential if $r \geq 2$. 

A possibly useful example is the Lorenz gauge, for which we set $\deltabf\lintprod\vp = 0$. In this case, the exterior-calculus identity~\eqref{eq:left-wedge} allows us to write the inhomogeneous Maxwell equation~\eqref{eq:maxwell-Gen1} as
\begin{gather} 
\deltabf\lintprod \mf = (-1)^{r-1}(\deltabf\cdot\deltabf)\vp = \sd.
\end{gather}
In this gauge, the Maxwell equations become $\binom{k+n}{r-1}$ uncoupled ultrahyperbolic wave equations for the separate components of $\sd$. Similarly, we may choose a transverse gauge where not only $\deltabf\lintprod\vp = 0$ is satisfied, but also $\deltabf_\text{t}\lintprod\vp = \deltabf_\text{s}\lintprod\vp = 0$, where $\deltabf_\text{t}$ and $\deltabf_\text{s}$ respectively represent the time and space components of $\deltabf$. 

Considering an $(r-2)$-vector gauge field $\gf$, the Lorenz gauge condition together with~\eqref{eq:left-wedge} imply that 
\begin{equation}\label{eq:gauge-harmonic}
	\deltabf\lintprod(\deltabf\wedge\gf ) = (-1)^{r-2}(\deltabf\cdot\deltabf)\gf + \deltabf\wedge(\deltabf\lintprod\gf) = 0.
\end{equation}
When $r = 2$, the interior derivative $\deltabf\lintprod\gf$ of a scalar gauge field $\gf$ vanishes and we find that $(\deltabf\cdot\deltabf)\gf = 0$, namely that the gauge field $\gf$ must be a harmonic function. For larger values of $r$, however, one should consider the full equation~\eqref{eq:gauge-harmonic}.

\subsection{Maxwell Equations in the Fourier Domain}
\label{subsec-fourier}

It is instructive to express the Maxwell equations in the Fourier domain, especially to study the field propagation in the absence of sources. Let the Fourier variable be denoted by $\xibf = (\xi_0,\dotsc,\xi_{k+n-1})$ and the Fourier transform $\mff(\xibf)$ of the Maxwell field be
\begin{equation}\label{eq:F-fourier}
	\mff(\xibf) = \idotsint \drm^{k+n}\xb \,e^{-j2\pi\xibf\cdot\xb}\mf(\xb).
\end{equation}
We also have a similar expression for the Fourier transform $\sdf(\xibf)$ of the source.
The Maxwell equations~\eqref{eq:maxwell-Gen1} and~\eqref{eq:maxwell-Gen2} adopt the algebraic form
\begin{gather} 
j2\pi\xibf\lintprod \mff = \sdf, \\
\xibf\wedge \mff = 0. 
\end{gather}
For the Fourier transform of the vector potential $\vpf(\xibf)$, the identity $\mf = \deltabf\wedge\vp$ implies that $\mff(\xibf) = j2\pi\xibf\wedge\vpf(\xibf)$.

In the absence of sources, the Maxwell equations~\eqref{eq:maxwell-Gen1} and~\eqref{eq:maxwell-Gen2} adopt the simpler form
\begin{gather} 
\xibf\lintprod \mff = 0, \\
\xibf\wedge \mff = 0. 
\end{gather}
The first equation would seem to require that $\mff$ is orthogonal to $\xibf$, while the second requires that $\mff$ is parallel to $\xibf$. The only non-trivial combination of these two possibilities is that $\xibf$ has zero norm, that is $\xibf\cdot\xibf = 0$. This is possible only if both $k$ and $n$ are positive numbers. 
Using~\eqref{eq:left-wedge} and these two Maxwell equations in the Fourier domain, we have
\begin{equation}
	0 = \xibf\lintprod(\xibf\wedge\mff) = (-1)^{r}(\xibf\cdot\xibf)\mff + \xibf\wedge(\xibf\lintprod\mff) = (-1)^{r-1}(\xibf\cdot\xibf)\mff,
\end{equation}
from which we conclude that the Fourier transform of the fields is supported only in the set $\xibf\cdot\xibf = 0$. With some abuse of notation, let $\mff$ denote this function with support only in the set $\xibf\cdot\xibf = 0$. 
We postulate that the inverse Fourier transform of the fields in free space is indeed given by
\begin{equation}	\label{eq:fourtransF}
	\mf(\xb) = \idotsint \drm^{k+n}\xibf \,e^{j2\pi\xibf\cdot\xb}\mff(\xibf)\delta(\xibf\cdot\xibf).
\end{equation}

\subsection{Integral Form of Maxwell Equations}

The generalized Maxwell equations also admit an integral form obtained by using the Stokes theorem \cite[Sect.~36.D]{arnold1989mathematicalMethods}. Moreover, these integral expressions are associated with two natural operations on the Maxwell field, namely its circulation and its flux. These associations are  particularly transparent in the exterior-algebra formulation presented in this paper, while being simultaneously valid for generic values of $r$, $k$, and $n$. From~\eqref{eq:circulation} and~\eqref{eq:maxwell-Gen2}, and noting that the interior product is now a dot product, we find that the circulation of the Maxwell field $\Fb$ along the boundary of any $(r+1)$-dimensional space-time volume $\mathcal{V}^{r+1}$ is zero:
\begin{align}
\int_{\partial\mathcal{V}^{r+1}}\drm^{r}\xb\cdot\Fb &= \int_{\mathcal{V}^{r+1}}\drm^{r+1}\xb\cdot(\deltabf\wedge \Fb) \\ &= 0.\label{eq:86}
\end{align}
For $r = 2$, $k = 1$ and $n = 3$,~\eqref{eq:86} is a scalar equation and we obtain the usual integral forms of the pair of homogeneous Maxwell equations by considering two different hypersurfaces, all-space, and time-space, as verified in detail in~\cite[Sect.~4.3]{colombaro2019introductionSpaceTimeExteriorCalculus}.

Similarly, from~\eqref{eq:flux-2} and~\eqref{eq:maxwell-Gen1}, the flux of the Maxwell field $\Fb$ across the boundary of any $(k+n-r+1)$-dimensional space-time volume is equal to the flux of the current density $\Jb$ across the $(k+n-r+1)$-dimensional space-time volume:
\begin{align}
\int_{\partial\mathcal{V}^{k+n-r+1}}\drm^{k+n-r}\xb^{\hodgeinv}\cdot\Fb &= 
	\int_{\mathcal{V}^{k+n-r+1}}\drm^{k+n-r}\xb^{\hodgeinv}\cdot(\deltabf\lintprod \Fb) \\ &= \int_{\mathcal{V}^{k+n-r+1}}\drm^{k+n-r+1}\xb^{\hodgeinv}\cdot\Jb.\label{eq:93}
\end{align}	
As with the homogeneous Maxwell equations, for $r = 2$, $k = 1$, and $n = 3$ the scalar equation~\eqref{eq:93} yields the usual integral inhomogeneous equations by considering two different hypersurfaces $\mathcal{V}^{3}$ \cite[Sect.~4.3]{colombaro2019introductionSpaceTimeExteriorCalculus}, respectively all-space and time-space.

\section{Lorentz Force and Stress-Energy-Momentum Tensor}
\label{sec:lorentz}
As stated in the Introduction, the action of the electromagnetic field on the charges is described by the Lorentz force. In relativistic form and tensor notation, the four-dimensional Lorentz force density vector $\fb$ is related to the electromagnetic field $F^{\alpha\beta}$ and the source density $J^\beta$ as given in~\eqref{eq:lorentz-force-tens},
\begin{equation}\label{eq:lorentz-force-tens-bis}
	\text{f}^{\alpha} = F^{\alpha\beta}J_\beta.
\end{equation} 
The interaction between fields and charges involves a transfer of energy and momentum between the former and the latter.
This interaction is subject to a conservation law relating the Lorentz force density $\text{f}^\alpha$ and the (symmetric) stress-energy-momentum tensor $T^{\alpha\beta}$ of the electromagnetic field \cite[Sect.~32]{landau1982classicalTheoryFields} or \cite[Sect.~12.10]{jackson1999classicalElectrodynamics}, 
\begin{equation} \label{eq:lorentz-energy-tensor}
	\text{f}^{\alpha} +\partial_\beta T^{\alpha\beta} = 0.
\end{equation} 
For any values of $k$ and $n$ and irrespective of the value of $r$, both the Maxwell field and the source density carry energy-momentum.
This generalized energy-momentum is a vector with $k+n$ components, the first $k$ (resp.~remaining $n$) of which represent temporal (resp.~spatial) components. 

As we discuss in Sect.~\ref{subsec:lorentz-density}, the generalized Lorentz force density remains a $1$-vector with $k+n$ components. Integrated over a space-time region, this density characterizes the action of the field upon the source density.
As explained in Sect.~\ref{subsec:stress-energy-momentum}, this interaction is also subject to a conservation law similar to~\eqref{eq:lorentz-energy-tensor}.
The stress-energy-momentum tensor of the Maxwell field is found to be a symmetric bitensor of rank 2 for any values of $r$, $k$ and $n$.
Finally, as an application, Sect.~\ref{subsec:flux-SEM-tensor} studies the flux of the stress-energy-momentum tensor across a $(k+n-1)$-dimensional slice of space-time and gives an expression for this flux in terms of the Fourier transform of the potential.


\subsection{Lorentz Force Density}
\label{subsec:lorentz-density}

Energy-momentum can be transferred from the fields to the charges through a process modelled as a force acting on the charges.
We refer to the Lorentz force, whose density was introduced in~\eqref{eq:lorentz-force-tens}. 
In exterior-calculus form, the generalized Lorentz force density $\fb$ is a vector of grade 1 with $k+n$ components given by 
\begin{equation}\label{eq:lorentz}
	\fb = \Jb\lintprod \Fb = (\deltabf\lintprod \mf)\lintprod \mf.
\end{equation}
The volume integral of the Lorentz force density $\fb$ over an $(k+n)$-dimensional hypervolume $\mathcal{V}^{k+n}$ quantifies the transfer of energy-momentum to the charges in that volume. In turn, from the discussion in Sect.~2.2, this volume integral is the flux of $\fb$ over the volume.  As the grade of $\Fb$ is $r$ and that of $\Jb$ is $r-1$, and the left interior product lowers the grade, the result has indeed grade $r-(r-1) = 1$. It is rather straightforward to verify that this Lorentz force coincides with the relativistic Lorentz force for $r = 2$, $k = 1$ and $n = 3$ \cite[Sect.~4.1]{colombaro2019introductionSpaceTimeExteriorCalculus}, and with the forces upon charges or currents in electrostatics or magnetostatics.

\subsection{Stress-Energy-Momentum Tensor}
\label{subsec:stress-energy-momentum}

The flux or transfer of energy-momentum over space-time is described by means of the stress-energy-momentum tensor $\Tb(\xb)$, a bitensor with $\smash{\binom{k+n+1}{2}}$ independent components at each point of space-time, regardless of the value of the grade $r$,  
\begin{equation}\label{eq:T_Tij}
	\Tb = \sum_{i\leq j}T_{ij}\ubf_{ij}.
\end{equation}
Here, the bitensor basis elements are denoted by $\smash{\ubf_{ij}}$, where $(i,j)$ is an ordered list of two (possibly repeated) space and time indices. 
In the usual tensor algebra, this bitensor would be a symmetric tensor of rank 2 as we can identify $\ubf_{ij}$ as $\ubf_{ij} = \ebf_i\vee\ebf_j$, the symmetric tensor product of $\ebf_i$ and $\ebf_j$. 
A benefit of using exterior algebra is that the stress-energy-momentum tensor of the Maxwell field admits a rather simple formula valid for any triplet $r$, $k$, and $n$. 

Before proceeding further, we define the left and right interior product of a vector $\ebf_{i}$ and a bitensor $\ubf_{jk}$, two bilinear operations giving a vector. These products are the extension by linearity of the product of unit basis vectors given by
\begin{equation}\label{eq:int-prod-bitensor}
\ebf_{i}\lintprod \ubf_{jk} = \ubf_{jk} \rintprod \ebf_{i} = \begin{cases} \ebf_{j}\Delta_{ii}, \quad &i=k, \\ \ebf_{k}\Delta_{ii}, \quad &i=j,\\ 0, &\mathrm{otherwise}.\end{cases}
\end{equation}
Let $\deltabf\lintprod$ denote the interior derivative, as in~\eqref{eq:int-prod-bitensor}. We prove in~\ref{app:Tij} the conservation law for the equivalent energy-momentum relating the Lorentz force~\eqref{eq:lorentz} and the stress-energy-momentum tensor $\Tb$ of the Maxwell field $\mf$
\begin{equation}\label{eq:conservation-EM}
	\fb + \deltabf\lintprod \Tb = 0.
\end{equation}	
Here, the stress-energy-momentum tensor $\Tb$ of the Maxwell field $\mf$ is given by the expression
\begin{equation}\label{eq:totalT}
\Tb = -(\mf\odot\mf+\mf\owedge\mf),
\end{equation}
where $\mf\odot\mf$ and $\mf\owedge\mf$ are two bitensors with components respectively given by
\begin{align}
 \mf\odot\mf\bigr|_{ij} &= \frac{1}{2}\Delta_{ii}\Delta_{jj}(\ebf_{i}\lintprod \mf)\cdot (\mf\rintprod \ebf_{j}) \label{eq:Tij-int} \\
 \mf\owedge\mf\bigr|_{ij} &= \frac{1}{2}\Delta_{ii}\Delta_{jj}(\ebf_{i}\wedge\mf)\cdot(\mf \wedge \ebf_{j}). \label{eq:Tij-ext}
\end{align}

In~\ref{app:components} we compute the explicit expressions for $T_{ij}$, 
namely
\begin{align}
	T_{ii} &= \frac{(-1)^{r}}{2}\Delta_{ii}\Biggl(\sum_{I\in\mathcal{I}_{r}: i \in I}\mfi_I^2\Delta_{II} - \sum_{I\in\mathcal{I}_{r}: i \notin I}\mfi_{I}^2\Delta_{II}\Biggr) \label{eq:Tii_final} \\
	T_{ij} &= -\sum_{L\in\mathcal{I}_{r-1}}\sigma(L,i)\sigma(j,L)\mfi_{i+L} \mfi_{j+L} \Delta_{LL}. \label{eq:Tij_final}
\end{align}
A tedious calculation would serve to verify that the tensor components in~\eqref{eq:Tij-int} and~\eqref{eq:Tij-ext} indeed coincide with the well-known electromagnetic stress-energy tensor for $r = 2$, $k = 1$, and $n = 3$ \cite[Sect.~32]{landau1982classicalTheoryFields}. 
In general, the components of $\Tb$ come in two different forms. For $i = j$, the component $T_{ii}$ in~\eqref{eq:Tii_final} is given by the sum of the squares of all $\binom{k+n}{r}$ components in the Maxwell field $F_I$, each weighted by either $+\frac{1}{2}$ or $-\frac{1}{2}$ depending of the values of $i$ and $I$. For $i \neq j$, the component $T_{ij}$ in~\eqref{eq:Tij_final} is given by the sum of $\binom{k+n-2}{r-1}$ terms, each of this is a product of two different components of the Maxwell field weighted by either +1 or -1. 

The values of $T_{ij}$ in~\eqref{eq:Tij-int} and~\eqref{eq:Tij-ext} do not change if we swap the values of $i$ and $j$ so $\Tb$ indeed corresponds to a symmetric tensor. To verify this fact, we note that from~\eqref{eq:comm-intprod}, it holds that $\ebf_{j}\lintprod \Fb = (-1)^{r+1}\Fb\rintprod \ebf_{j}$. This implies that
\begin{align}
	(\ebf_{j}\lintprod \Fb )\cdot (\Fb\rintprod \ebf_{i}) &=  (\Fb\rintprod \ebf_{i}) \cdot (\ebf_{j}\lintprod \Fb ) \\
	&= (\ebf_{i}\lintprod \Fb )(-1)^{r+1} \cdot (\Fb\rintprod \ebf_{j})(-1)^{r+1} \\
	&= (\ebf_{i}\lintprod \Fb )\cdot (\Fb\rintprod \ebf_{j}).
\end{align}
Therefore the components of $\Fb\odot\Fb$ in~\eqref{eq:Tij-int} are symmetric. 
Concerning the components of $\Fb\wedge\Fb$ in~\eqref{eq:Tij-ext}, one proves that they are symmetric by using a similar reasoning exploiting that $\ebf_{j}\wedge\Fb = (-1)^r (\Fb\wedge \ebf_{j})$ from~\eqref{eq:comm-extprod}.

The tensor trace $\Tr \Tb = \sum_i \Delta_{ii}T_{ii}$ is given for a generic triplet $k$, $n$, and $r$ by
\begin{align}
 \Tr \Tb &= \frac{(-1)^{r}}{2}\sum_i \Delta_{ii}^2 \Biggl(\sum_{I\in\mathcal{I}_{r}: i \in I}\mfi_I^2\Delta_{II} - \sum_{I\in\mathcal{I}_{r}: i \notin I}\mfi_{I}^2\Delta_{II}\Biggr) \label{eq:A54}
 \\
 &= \frac{(-1)^{r}}{2}\sum_{I\in\mathcal{I}_{r}}\mfi_I^2\Delta_{II} \Biggl(\sum_{i: i \in I}1 - \sum_{i: i \notin I}1\Biggr) \label{eq:A55}
 \\
 &= \frac{(-1)^{r+1}}{2}(k+n-2r)\,\mf\cdot\mf \label{eq:A56},
\end{align}
where we swapped the summation order in~\eqref{eq:A55} and extracted a common factor and then we used that $ \Fb \cdot \Fb = \sum_I \mfi_I^2 \Delta_{II}$ and that $\sum_{i: i \in I} 1 = r$ and that $\sum_{i: i \not\in I} 1 = n+k-r$ in~\eqref{eq:A56}.
The tensor is traceless for $k+n = 2r$ or in the case that $\mf\cdot\mf = 0$. To any extent, the trace is a Lorentz invariant.

\subsection{Flux of the Stress-Energy-Momentum Tensor}
\label{subsec:flux-SEM-tensor}

The energy-momentum conservation law also admits an integral form, which we derive next. First, the volume integral of the Lorentz force density $\fb$ over an $(k+n)$-dimensional hypervolume $\mathcal{V}^{k+n}$ gives the transfer of energy-momentum to the charges in that volume. From the discussion in Sect.~2.2, this volume integral is the flux of $\fb$ over $\mathcal{V}^{k+n}$, and from the conservation law in differential form~\eqref{eq:conservation-EM}, we obtain the following integral representation of the energy-momentum transfer
\begin{equation}
	\int_{\mathcal{V}^{k+n}}\fb\drm x_{0,\dotsm,{k+n-1}} = \int_{\mathcal{V}^{k+n}}\drm^{k+n}\xb^{\hodgeinv}\lintprod\fb = -\int_{\mathcal{V}^{k+n}}\drm^{k+n}\xb^{\hodgeinv}\lintprod(\deltabf\lintprod\Tb).
\end{equation}
A short extension to the proof of~\eqref{eq:flux-2} in~\cite[Sect.~3.5]{colombaro2019introductionSpaceTimeExteriorCalculus}, included in~\ref{app:stokes_bitensor} proves a Stokes theorem for bitensors: 
the flux of a bitensor field $\Tb$ across the boundary $\partial\mathcal{V}^{\ell}$ of an $\ell$-dimensional hypersurface $\mathcal{V}^{\ell}$ is equal to the flux of the interior derivative of $\Tb$ across $\mathcal{V}^{\ell}$ for any $\ell$, and in particular for $\ell = k+n$. Using this form of the Stokes theorem, we obtain 
%
\begin{equation}\label{eq:flux-T}
	\int_{\mathcal{V}^{k+n}}\fb\drm x_{0,\dotsm,{k+n-1}} = -\int_{\partial\mathcal{V}^{k+n}}\drm^{k+n-1}\xb^{\hodgeinv}\lintprod\Tb.
\end{equation}

As an example, and for some fixed $x_\ell$ and $\ell \in\{0,\dotsc,k+n-1\}$, consider a $(k+n)$-dimensional space-time region of the form
\begin{equation} \label{eq:integr-region}
	\mathcal{V}_\ell^{k+n} = (-\infty,\infty)\times(-\infty,\infty)\dotsm \times (-\infty,x_\ell)\times\dotsm(-\infty,\infty).
\end{equation}
This region is a half space-time with boundary a surface with constant space-time coordinate $\ell$ of value $x_\ell$ given by
\begin{equation} \label{eq:integr-region-boundary}
	\partial\mathcal{V}_\ell^{k+n} = (-\infty,\infty)\times(-\infty,\infty)\dotsm \times \{x_\ell\}\times\dotsm(-\infty,\infty).
\end{equation}
In the computation of the flux of $\Tb$ across the boundary $\partial\mathcal{V}^{k+n}$ in~\eqref{eq:flux-T}, the infinitesimal vector element is given by
\begin{equation}
	\drm^{k+n-1}\xb^\hodgeinv = \drm x_{\ell^c}\sigma(\ell,\ell^c)\Delta_{\ell\ell}\ebf_{\ell},
\end{equation}	
where the proper orientation carries an additional factor $\sigma(\ell,\ell^c)$, the normal vector pointing outside the integration region being $\ebf_\ell$. Using~\eqref{eq:T_Tij}, the flux is an integration over all space-time dimensions other than the $\ell$-th,
\begin{align}
	\int_{\partial\mathcal{V}_\ell^{k+n}}\drm^{k+n-1}\xb^\hodgeinv \lintprod \Tb(\xb) &= \int_{-\infty}^\infty\dotsm\int_{-\infty}^\infty \drm x_{\ell^c}\sigma(\ell,\ell^c)\Delta_{\ell\ell}\ebf_{\ell} \lintprod\Tb(\xb) \label{eq:flux_T} \\
	&= \sum_{i=0}^{k+n-1}\ebf_{i}\int_{-\infty}^\infty\dotsm\int_{-\infty}^\infty \drm x_{\ell^c}\sigma(\ell,\ell^c) T_{i\ell}(\xb),\label{eq:flux_Til}
\end{align}	
where we used the symmetry of $T_{ij}$ between $i$ and $j$ and the interior product in~\eqref{eq:int-prod-bitensor}. This formula represents the fact that the component $T_{i\ell}(\xb)$ characterizes the flux of the $i$-th component of the energy-momentum vector across a surface with constant space-time coordinate $\ell$, namely the boundary of $\mathcal{V}_\ell^{k+n}$.
For instance, for $r=2$, $k = 1$, $n = 3$, and $\ell = 0$ the integrals in~\eqref{eq:flux_Til} give the energy (for $i = 0$) and the momentum (for $i = 1, 2, 3$), in line with the fact that $T_{00}(\xb)$ represents the energy density $U$ and $T_{0i}(\xb)$ the three components of the Poynting vector $\Sb$. 
In this case, the time component of~\eqref{eq:conservation-EM} gives the standard conservation law $\partial_0 U + \nabla\cdot \Sb = -f_0$, $f_0$ being the work done on the charges by the field.

The flux~\eqref{eq:flux_Til} across the surface with constant space-time coordinate $\ell$ can also be expressed in a relatively compact form that involves the Fourier transform of the vector potential $\vpf(\xibf)$ in the Lorenz gauge introduced in Sect.~\ref{sec:vector_potential}. As we prove in~\ref{app:flux-T}, this flux is given by
\begin{equation}
 \label{eq:flux-fixed-ell}
\int_{\partial\mathcal{V}_\ell^{k+n}}\drm^{k+n-1}\xb^\hodgeinv \lintprod \Tb 
= (-1)^{r}2\pi^2\sigma (\ell , \ell^c) 
\int_{\Delta_{\ell\ell} \xibf_{\bar\ell} \cdot \xibf_{\bar\ell}\le 0}
\drm \xi_{\ell^c} 
 \frac{\xibf_{+}}{\xi_{+,\ell}}\bigl\vert \hat{\Ab}(\xibf_+)\bigr\vert^2,
\end{equation}
where $\xibf_+$ are frequency vectors satisfying $\xibf_+\cdot\xibf_+ = 0$ given by
\begin{align}
\xibf_+ &=(\xi_0, \dotsc, \xi_{\ell-1},+\sqrt{-\Delta_{\ell\ell}   \xibf_{\bar\ell} \cdot \xibf_{\bar\ell}},  \xi_{\ell+1}, \dotsc, \xi_{k+n-1}),\\
\xibf_{\bar\ell} &=(\xi_0, \dotsc,\xi_{\ell-1},0,\xi_{\ell+1}, \dotsc, \xi_{k+n-1}).
\end{align}
This expression is reminiscent of the one appearing in the second quantization of the electromagnetic field \cite[Sect.~2]{berestetskii1982quantum}, where the Hamiltonian, i.~e.~$i = \ell = 0$, is picked as the starting point for the quantization procedure, and the independent degrees of freedom of the vector potential are quantized. 
Eq.~\eqref{eq:flux-fixed-ell} suggests that quantization of the vector potential does not require starting with the Hamiltonian. Having said this, a possible quantization of the generalized Maxwell equations would have to properly deal with the gauge independence of the theory. There exist several methods of deriving gauge-independent quantum Maxwell equations, such as path-integral and BRST quantizations. Studying the applicability of these methods to the exterior-algebra multivectors is beyond the scope of this work.

We conclude this section with a brief discussion on the number of degrees of freedom in the Maxwell field in the absence of sources.
Eq.~\eqref{eq:flux-fixed-ell} expresses the change of energy-momentum as a linear superposition of the squared modulus of the Fourier transform of the vector potential $\vpf(\xibf)$. If both $k$ and $n$ are positive, for any frequency vector $\xibf$, we may choose a transverse gauge where not only $\xibf\lintprod\vpf = 0$ is satisfied, but also $\xibf_\text{t}\lintprod\vpf = \xibf_\text{s}\lintprod\vpf = 0$, where $\xibf_\text{t}$ and $\xibf_\text{s}$ respectively represent the time and space components of $\xibf$. The Fourier vector potential $\vpf(\xibf)$ is thus perpendicular to the vectors $\xibf_\text{t}$ and $\xibf_\text{s}$, and the number of available dimensions of space-time where  $\vpf(\xibf)$ lies is reduced to a total of $k+n-2$. The number of independent components of $\vpf(\xibf)$, or degrees of freedom, is therefore $\binom{k+n-2}{r-1}$, e.~g.~two polarizations for $r=2$, $k = 1$, and $n = 3$. Moreover, these independent degrees of freedom have the form of propagating waves in space-time.


\section{Conclusions and Future Work}

In this paper, we have put forward exterior algebra and calculus as a natural setting for an abstract yet intuitively simple form of generalized Maxwell equations, Lorentz force density, and stress-energy tensor for generalized electromagnetic fields (or Maxwell fields) represented by multivectors of grade $r$ in a space-time with an arbitrary number of dimensions. The source density is modeled as a multivector of grade $r-1$. The phenomenological description of the charges associated to these source densities will be done elsewhere. 

The generalized Maxwell equations are given in terms of exterior-calculus operations. In differential form, the homogeneous Maxwell equation states that the exterior derivative of the Maxwell field is zero; the inhomogeneous Maxwell equation states that the interior derivative of the Maxwell field is the source density. In integral form, the homogeneous equation states that the circulation of the Maxwell field along the boundary of any $(r + 1)$-dimensional space-time volume is zero; the inhomogeneous equation states that the flux of the Maxwell field across the boundary of any $(k+n-r+1)$-dimensional space-time hypervolume is equal to the flux of the current density across the same hypervolume. The Lorentz force density is given by the left interior product of the source density and the Maxwell field. Moreover, a conservation law relates this Lorentz force density with the interior derivative of the stress-energy-momentum tensor of the Maxwell field. 

Among several applications, a simple expression for the flux of the stress-energy-momentum tensor across an slice of space-time with a constant coordinate is given in terms of the Fourier transform of the potentials. 
The description of Maxwell fields is classical, as energy and momentum carried by the fields are continuous rather than discrete, in contrast with the experimental observations in the ordinary space-time. There exist several methods of postulating gauge-independent quantum Maxwell equations, such as second or canonical quantization or path integral quantization. A study of the applicability of these methods to the generalized Maxwell equations is left open for future work.

\appendix

\section{Proofs Related to the Stress-Energy-Momentum Tensor}

\subsection{Distributive Properties of the Interior Product}
\label{app:properties}

\paragraph{Proof of~\eqref{eq:text-equiv-wedge-int}}
%
To prove this equation, we first expand the vectors and multivectors in the left-hand side of~\eqref{eq:text-equiv-wedge-int} in terms of their components to get
\begin{align}
\left( \vb \wedge \wb \right) \cdot \left( \wb' \wedge \vb' \right) &= \left(\sum _{i,I:i\not\in I} v_i w_I \sigma(i,I) \ebf_{i+I} \right) \cdot \left(\sum _{j,J:j\not\in J} v_j' w_J' \sigma(J,j) \ebf_{j+J} \right) \\
&= \sum_{\substack{i,I:i\not\in I \\ j,J:j\not\in J}}
v_i w_I v_j' w_J' \sigma(i,I) \sigma(J,j) \Delta _{i+I, j+J}.
\end{align}

At this point, we separate the cases $i=j$ and $i\ne j$, namely
\begin{align} \label{eq:proof-mid-wed-int}
\left( \vb \wedge \wb \right) \cdot \left( \wb' \wedge \vb' \right) &= \sum _{i,I} v_i v_i' w_I w_I' \sigma(i,I) \sigma(I,i) \Delta_{ii} \Delta_{ii}
\notag	\\
&+ \sum _{i\ne j, I,J} v_i v_j' w_I w_J' \sigma(i,I) \sigma(J,j) \Delta_{ii} \Delta_{jj} \Delta_{I\setminus j,J \setminus i},
\end{align}
and it is easy to note that the first term can be written as
\begin{align}
 \sum _{i,I} v_i v_i' w_I w_I' \sigma^2(i,I) (-1)^r \Delta_{ii} \Delta_{ii} = (-1)^r  \left( \vb \cdot \vb' \right) \left( \wb \cdot \wb' \right).
\end{align}

Regarding the second term, we first prove the equality
\begin{equation} \label{eq:eq-proof-wedge-int}
 \sigma(i,I) \sigma(J,j)=\sigma(I\setminus j, j) \sigma(i, J\setminus i) ,
\end{equation}
which we can be rewritten in the form
\begin{equation} \label{eq:eq-proof-wedge-int2}
\sigma(I\setminus j, j) \sigma(i,I) = \sigma(i, J\setminus i) \sigma(J,j) .
\end{equation}

\begin{figure}[!htb]
\centering
	\begin{subfigure}[t]{0.45\textwidth}
	\centering
\begin{tikzpicture}
\draw [blue, thick] (-3,4) node[black] {$\vert$} -- (1,4) node[black] {$\vert$};
\node[blue, above] at (-1,4) {$I\setminus j$};
\draw [thick] (-3.5,4) node[black] {$\vert$} -- (-3,4) node[black] {$\vert$};
\node[above] at (-3.25,4) {$i$};
\draw [red, thick] (1,4) -- (1.5,4) node[black] {$\vert$};
\node[red, above] at (1.25,4) {$j$};
\draw [blue, thick] (-3,3)node[black] {$\vert$} -- (1.5,3) node[black] {$\vert$};
\node[blue, above] at (-0.75,3) {$I$};
\draw [thick] (-3.5,3) node[black] {$\vert$} -- (-3,3) node[black] {$\vert$};
\node[above] at (-3.25,3) {$i$};
\draw [thick] (-3.5,2)node[black] {$\vert$} -- (1.5,2) node[black] {$\vert$};
\node[above] at (-1,2) {$i+I$};
\end{tikzpicture}
\end{subfigure}
\centering
	\begin{subfigure}[t]{0.45\textwidth}
	\centering
\begin{tikzpicture}
\draw [blue, thick] (-3,4) node[black] {$\vert$} -- (1,4) node[black] {$\vert$};
\node[blue, above] at (-1,4) {$I\setminus j$};
\draw [thick] (-3.5,4) node[black] {$\vert$} -- (-3,4) node[black] {$\vert$};
\node[above] at (-3.25,4) {$i$};
\draw [red, thick] (1,4) -- (1.5,4) node[black] {$\vert$};
\node[red, above] at (1.25,4) {$j$};
\draw [thick]  (-3.5,3)node[black] {$\vert$} -- (1,3) node[black] {$\vert$};
\node[ above] at (-1.25,3) {$I\setminus i +j$};
\draw [red, thick] (1,3) -- (1.5,3) node[black] {$\vert$};
\node[red, above] at (1.25,3) {$j$};
\draw [thick] (-3.5,2)node[black] {$\vert$} -- (1.5,2) node[black] {$\vert$};
\node[above] at (-1,2) {$i+I$};
\end{tikzpicture}
\end{subfigure}
\caption{Visual aid for the identity among permutations in~Equation \eqref{eq:eq-proof-wedge-int2}.}
\label{fig:proof-wedge-int}
\end{figure}
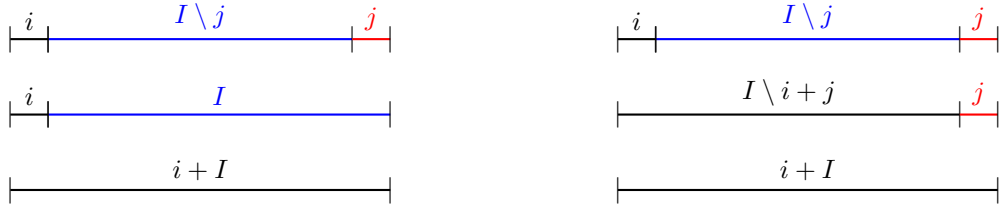

In fact, with the visual help of \figurename~\ref{fig:proof-wedge-int} it is easy to see how the sign of the permutation on the left-hand side of \eqref{eq:eq-proof-wedge-int2} is the same as that on the right-hand side.
Then, the second term of the expression \eqref{eq:proof-mid-wed-int} can be written
\begin{equation}
\sum_{i\ne j, I, J} v_j' w_I \sigma(I\setminus j,j) \Delta_{jj} \, v_i w_J' \sigma(i, J\setminus i) \Delta_{ii}
\, \Delta_{I\setminus j, J\setminus i} = (\vb'\lintprod \wb)\cdot (\wb'\rintprod\vb).
\end{equation}
\qed

\paragraph{Proof of~\eqref{eq:text-equiv-double-interior}} 
To prove this identity we write separately the explicit expressions of the left and the right side of~\eqref{eq:text-equiv-double-interior}. On the left side we get
\begin{align}
\ub \lintprod (\wb \rintprod \vb) &= \left(\sum_a u_a \ebf _a \right) \lintprod \left(\sum_{b\in I} v_b w_I \Delta_{b,b} \sigma(b,I\setminus b) \ebf_{I\setminus b} \right) \\
&= \sum_{\substack{a,b\in I\\a\ne b}} u_a v_b w_I \Delta_{a,a} \Delta_{b,b} \sigma(b,I\setminus b) \sigma(I\setminus b\setminus a, a) \ebf_{I\setminus b\setminus a}
, \label{eq:proof-intint-lhs}
\end{align}
while on the right side we result is
\begin{align}
(\ub \lintprod \wb) \rintprod \vb &= \left(\sum_{a\in I} u_a w_I \Delta_{a,a} \sigma(I\setminus a,a) \ebf _{I\setminus a} \right) \rintprod \left(\sum_{b} v_b \ebf_{b} \right) \\
&=
\sum_{\substack{a,b\in I\\a\ne b}} u_a v_b w_I \Delta_{a,a} \Delta_{b,b} \sigma(I\setminus a,a) \sigma(b, I\setminus a\setminus b) \ebf_{I\setminus a\setminus b}
. \label{eq:proof-intint-rhs}
\end{align}

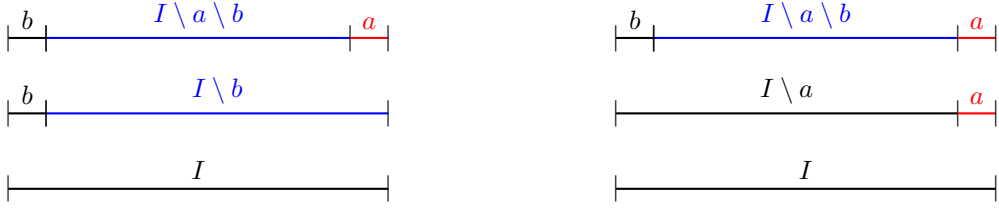
\begin{figure}[htb]
\centering
	\begin{subfigure}[t]{0.45\textwidth}
	\centering
\begin{tikzpicture}
\draw [blue, thick] (-3,4) node[black] {$\vert$} -- (1,4) node[black] {$\vert$};
\node[blue, above] at (-1,4) {$I\setminus a\setminus b$};
\draw [thick] (-3.5,4) node[black] {$\vert$} -- (-3,4) node[black] {$\vert$};
\node[above] at (-3.25,4) {$b$};
\draw [red, thick] (1,4) -- (1.5,4) node[black] {$\vert$};
\node[red, above] at (1.25,4) {$a$};
\draw [blue, thick] (-3,3)node[black] {$\vert$} -- (1.5,3) node[black] {$\vert$};
\node[blue, above] at (-0.75,3) {$I\setminus b$};
\draw [thick] (-3.5,3) node[black] {$\vert$} -- (-3,3) node[black] {$\vert$};
\node[above] at (-3.25,3) {$b$};
\draw [thick] (-3.5,2)node[black] {$\vert$} -- (1.5,2) node[black] {$\vert$};
\node[above] at (-1,2) {$I$};
\end{tikzpicture}
\end{subfigure}
\centering
	\begin{subfigure}[t]{0.45\textwidth}
	\centering
\begin{tikzpicture}
\draw [blue, thick] (-3,4) node[black] {$\vert$} -- (1,4) node[black] {$\vert$};
\node[blue, above] at (-1,4) {$I\setminus a\setminus b$};
\draw [thick] (-3.5,4) node[black] {$\vert$} -- (-3,4) node[black] {$\vert$};
\node[above] at (-3.25,4) {$b$};
\draw [red, thick] (1,4) -- (1.5,4) node[black] {$\vert$};
\node[red, above] at (1.25,4) {$a$};
\draw [thick]  (-3.5,3)node[black] {$\vert$} -- (1,3) node[black] {$\vert$};
\node[ above] at (-1.25,3) {$I\setminus a$};
\draw [red, thick] (1,3) -- (1.5,3) node[black] {$\vert$};
\node[red, above] at (1.25,3) {$a$};
\draw [thick] (-3.5,2)node[black] {$\vert$} -- (1.5,2) node[black] {$\vert$};
\node[above] at (-1,2) {$I$};
\end{tikzpicture}
\end{subfigure}
\caption{Visual aid for the identity among permutations in~Equation \eqref{eq:proof-int-int}.}
\label{fig:proof-int-int}
\end{figure}

For the expressions \eqref{eq:proof-intint-lhs} and \eqref{eq:proof-intint-rhs} to be identical, it is sufficient that
\begin{equation} \label{eq:proof-int-int}
\sigma(I\setminus b\setminus a, a) \sigma(b,I\setminus b) =
\sigma(b, I\setminus a\setminus b) \sigma(I\setminus a,a) 
\end{equation}
is satisfied. With the aid of \figurename~\ref{fig:proof-int-int}, we  notice that both sides represent the signature of two possible permutations reordering the list $(b,I\setminus a\setminus b, a)$ into $I$. This proves~\eqref{eq:text-equiv-double-interior}.
\qed

%

\paragraph{Proof of~\eqref{eq:text-wedge-dot-}}
We prove this relation by first writing separately the three terms of the equation. On the left-hand side, we get
\begin{align}
(\ub \wedge \vb) \cdot \wb &= \left( \sum_{i,I} u_i v_I \sigma(i,I)\ebf_{i+I} \right) \cdot \left( \sum_J w_J \ebf_J \right)\\
&= \sum_{\substack{i,I,J\\i+I=J}} u_i v_I w_J \sigma(i,I) \Delta_ {i+I,J}.
\end{align}
The central term is
\begin{align}
\vb \cdot (\wb \rintprod \ub ) &= \left( \sum_I v_I \ebf_I \right) \cdot \left( \sum_{i,J} u_i w_I \sigma(i,J\setminus i)\Delta_{ii}\ebf_{J\setminus i} \right) \\
&= \sum_{\substack{i,I,J\\J\setminus i=I}} u_i v_I w_J \sigma(i,I) \Delta_ {ii}\Delta_ {J\setminus i,I}.
\end{align}
Thus, it is easy to check that $\Delta_{i+I,J}=\Delta_ {ii}\Delta_ {J\setminus i,I}$, so that the first two terms of~\eqref{eq:text-wedge-dot-} coincide. Regarding the third term on the left hand side, we obtain
\begin{align}
\ub \cdot (\vb \lintprod \wb ) &= \left( \sum_i u_i \ebf_i\right) \cdot \left( \sum_{I,J} v_I w_J \sigma(J\setminus I,I)\Delta_{II}\ebf_{J\setminus I} \right) \\
&= \sum_{\substack{i,I,J\\i=J\setminus I}} u_i v_I w_J \sigma(i,I) \Delta_{i, J\setminus I}\Delta_ {II},
\end{align}
which corresponds to the first two expressions since $\Delta_{i, J\setminus I}\Delta_ {II} = \Delta_{i+I,J}=\Delta_ {ii}\Delta_ {J\setminus i,I}$.
\qed

\subsection{Interior Derivative of the Tensor}
\label{app:Tij}

For the sake of compactness, we define the bitensors $\Tb_\odot = \mf\odot\mf$ and $\Tb_\owedge = \mf\owedge\mf$ to prove the following identities
\begin{align}
	\deltabf\lintprod \Tb_\odot &= (\deltabf\lintprod \Fb)\lintprod \Fb \label{eq:delta-T1}\\
 	\deltabf\lintprod \Tb_\owedge &= (\deltabf\wedge\Fb) \rintprod \Fb. \label{eq:delta-T2}
\end{align}
Using equations~\eqref{eq:dot_multi},~\eqref{eq:ext-prod-def},~\eqref{eq:left-int-prod} and~\eqref{eq:right-int-prod}, we write $\Tb_\odot$ and $\Tb_\owedge $ explicitly in terms of components. That is, we obtain
\begin{align}
\Tb_\odot &=\frac{1}{2}  \sum _{i\le j} \Delta_{ii}\Delta_{jj}(\ebf_{i}\lintprod \Fb )\cdot (\Fb\rintprod \ebf_{j}) \, \ubf_{ij} \\
&=  \frac{1}{2} \sum _{i\le j} \sum_{\substack{I,J\in\mathcal{I}_r}}
\mfi_I \mfi_J \, \sigma(I\setminus i, i) \sigma(j, J\setminus j) \Delta_{I\setminus i, J\setminus j}
  \, \ubf_{ij}
\end{align}
and
\begin{align}
\Tb_\owedge &=\frac{1}{2}  \sum _{i\le j} \Delta_{ii}\Delta_{jj}(\ebf_{i}\wedge \Fb )\cdot (\Fb\wedge \ebf_{j}) \, \ubf_{ij} \\
&=  \frac{1}{2} \sum _{i\le j} \sum_{\substack{I,J\in\mathcal{I}_r}}
\mfi_I \mfi_J \, \sigma(i,I) \sigma(J,j)  \Delta_{ii}\Delta_{jj} \Delta_{I+ i, J+ j}
  \, \ubf_{ij}.
\end{align}

We start by computing the interior derivative $\deltabf \lintprod \Tb_\odot$ given by
\begin{multline}
\deltabf \lintprod \Tb_\odot =  \frac{1}{2} \sum_u \sum _{i\le j} \sum_{\substack{I,J}}
(\mfi_I \partial_u \mfi_J  + \mfi_J \partial_u \mfi_I) \\ \sigma(I\setminus i, i) \sigma(j, J\setminus j) \Delta_{I\setminus i, J\setminus j} \Delta_{uu}
  \, \ebf_u \lintprod \ubf_{ij}.
  \label{eq:A7}
\end{multline}
After some mathematical manipulations, equation~\eqref{eq:A7} is expanded as
\begin{align}
\deltabf \lintprod \Tb_\odot &= \sum_{i=j} \sum_{\substack{I,J}} \mfi_I \partial_i \mfi_J \sigma (I\setminus i,i) \sigma (j,J\setminus j) \Delta_{I\setminus i, J\setminus j} \, \ebf_i\phantom{.}\notag \\
  &\quad+ \frac{1}{2}\sum_{i<j} \sum_{\substack{I,J}} (\mfi_I \partial_j \mfi_J +\mfi_J \partial_j \mfi_I) \sigma (I\setminus i,i) \sigma (j,J\setminus j) \Delta_{I\setminus i, J\setminus j} \, \ebf_i \notag \\
   &\quad+\frac{1}{2} \sum_{i<j} \sum_{\substack{I,J}} (\mfi_I \partial_i \mfi_J +\mfi_J \partial_i \mfi_I) \sigma (I\setminus i,i) \sigma (j,J\setminus j) \Delta_{I\setminus i, J\setminus j} \, \ebf_j.
   \label{eq:A8}
\end{align}
Since $\Delta_{I\setminus i, J\setminus j}$ is nonzero only if $J\setminus j = I\setminus i$, we can use this condition in the relation $\sigma (j,J\setminus j) = \sigma (j, I\setminus i) = \sigma (I\setminus i, j) (-1)^{r-1} $ such that~\eqref{eq:A8} becomes
\begin{align}
\deltabf \lintprod \Tb_\odot &= \sum_{i=j} \sum_{\substack{I,J}} \mfi_I \partial_i \mfi_J \sigma (I\setminus i,i)\sigma (I\setminus i, j) (-1)^{r-1}\Delta_{I\setminus i, J\setminus j} \, \ebf_i\notag \\
  &\quad+ \frac{1}{2}\sum_{i<j} \sum_{\substack{I,J}} (\mfi_I \partial_j \mfi_J +\mfi_J \partial_j \mfi_I) \sigma (I\setminus i,i)  \sigma (I\setminus i, j) (-1)^{r-1} \Delta_{I\setminus i, J\setminus j} \, \ebf_i\notag \\
   &\quad+ \frac{1}{2}\sum_{i>j} \sum_{\substack{I,J}} (\mfi_I \partial_j \mfi_J +\mfi_J \partial_j \mfi_I) \sigma (I\setminus i,i)  \sigma (I\setminus i, j) (-1)^{r-1} \Delta_{I\setminus i, J\setminus j} \, \ebf_i.
   \label{eq:A9}
\end{align}
Exchanging the indices $i$ and $j$ and the labels $I$ and $J$, we get the following simplified expression
\begin{equation}
\deltabf \lintprod \Tb_\odot   =\sum_{i,j} \sum_{\substack{I,J}} \mfi_I \partial _j \mfi_J \, \sigma(I\setminus i, i) \sigma(I\setminus i, j) \Delta_{I\setminus i, J\setminus j} (-1)^{r-1} \, \ebf _i .\label{eq:deltaT1}
\end{equation}

A similar reasoning can be followed for the operation $\deltabf \lintprod \Tb_\owedge$ given by
\begin{multline}
\deltabf \lintprod \Tb_\owedge  = \frac{1}{2} \sum_u \sum _{i\le j} \sum_{\substack{I,J}}
(\mfi_I \partial_u \mfi_J  + \mfi_J \partial_u \mfi_I) \, \sigma(i,I) \sigma(J,j) \\\Delta_{I+i, J+j} \Delta_{u,u} \Delta_{i,i}\Delta_{j,j}
  \, \ebf_u \lintprod \ubf_{ij}.
 \end{multline}
The former equation can be expanded as
 \begin{align}
\deltabf \lintprod \Tb_\owedge   &= \sum_{i=j} \sum_{\substack{I,J}} \mfi_I \partial_i \mfi_J \sigma(i,I) \sigma(J,j) \Delta_{I+i, J+j} \, \ebf_i \notag \\
  &\quad+ \frac{1}{2}\sum_{i<j} \sum_{\substack{I,J}} (\mfi_I \partial_j \mfi_J +\mfi_J \partial_j \mfi_I)\sigma(i,I) \sigma(J,j) \Delta_{I+i, J+j}\Delta_{i,i}\Delta_{j,j} \, \ebf_i \notag\\
   &\quad+ \frac{1}{2}\sum_{i<j} \sum_{\substack{I,J}} (\mfi_I \partial_i \mfi_J +\mfi_J \partial_i \mfi_I) \sigma(i,I) \sigma(J,j) \Delta_{I+i, J+j}\Delta_{i,i}\Delta_{j,j} \, \ebf_j,
   \label{eq:A12}
\end{align}
and after a few manipulations using the properties of the signatures as done to obtain \eqref{eq:A9}, from~\eqref{eq:A12} we have
\begin{align}
 \deltabf \lintprod \Tb_\owedge  &= \sum_{i=j} \sum_{\substack{I,J\\I+i = J+j}} \mfi_I \partial_i \mfi_J \sigma(i,I) \sigma(j,J) (-1)^{r} \Delta_{I, I}\Delta_{j, j} \, \ebf_i \notag\\
  &\quad+ \frac{1}{2}\sum_{i<j} \sum_{\substack{I,J\\I+ i = J+j}} (\mfi_I \partial_j \mfi_J +\mfi_J \partial_j \mfi_I)\sigma(i,I) \sigma(j,J) (-1)^{r} \Delta_{I, I}\Delta_{j,j} \, \ebf_i \notag\\
   &\quad+ \frac{1}{2}\sum_{i>j} \sum_{\substack{I,J\\I+ i = J+j}} (\mfi_I \partial_j \mfi_J +\mfi_J \partial_j \mfi_I) \sigma(i,I)\sigma(j,J) (-1)^{r} \Delta_{I, I}\Delta_{j,j} \, \ebf_i.
\end{align}
Simplifying terms, we obtain
\begin{equation}
 \deltabf \lintprod \Tb_\owedge    = \sum_{i,j} \sum_{\substack{I,J\\I+ i = J+j}} \mfi_I \partial _j \mfi_J \, \sigma(i,I) \sigma(j,J) \Delta_{I,I}\Delta_{j,j} (-1)^{r} \, \ebf _i.\label{eq:deltaT2}
\end{equation}

We next derive explicit forms for the operations $(\deltabf\lintprod \Fb)\lintprod \Fb$ and $ (\deltabf\wedge\Fb) \rintprod \Fb$. We start by noting that $\deltabf\lintprod \Fb$ can be expanded in terms of component as
\begin{align}
\deltabf\lintprod \Fb &=
\biggl(\sum_j \Delta_{j,j} \partial _j \ebf_j \biggr)\lintprod \biggl(\sum_B F_B \ebf_B\biggr) \\&= \sum_{\substack{B,j\\j\in B}} \partial_j F_B \sigma(B\setminus j,j) \, \ebf_{B\setminus j}.
\end{align}
As a consequence, we have that
\begin{align}
(\deltabf\lintprod \Fb )\lintprod\Fb = \sum_{\substack{A,B,j\\(B\setminus j)\in A}} F_A ( \partial_j F_B ) \sigma(B\setminus j,j) \Delta_{B\setminus j,B\setminus j} \sigma(A\setminus (B\setminus j),B\setminus j)
\, \ebf_{A \setminus (B\setminus j)}.
\end{align}
Noting that $A \setminus (B\setminus j)$ is a single element $i$ such that $A\setminus i = B\setminus j$, we finally get
\begin{align}
(\deltabf\lintprod \Fb )\lintprod\Fb &= \sum_{i,j} \sum_{\substack{A,B\\A\setminus i = B\setminus j}} F_A ( \partial_j F_B ) \sigma(A\setminus i,j)  \sigma(i ,A\setminus i) \Delta_{A\setminus i,A\setminus i} 
\, \ebf_{i} \\
&= \sum_{i,j} \sum_{\substack{A,B\\A\setminus i = B\setminus j}} F_A ( \partial_j F_B ) \sigma(A\setminus i,j)  \sigma(A\setminus i,i) (-1)^{r-1} \Delta_{A\setminus i,A\setminus i} 
\, \ebf_{i} .
\label{eq:A19}
\end{align}
Since~\eqref{eq:A19} corresponds exactly to \eqref{eq:deltaT1}, we proved~\eqref{eq:delta-T1}.

Similarly, we next write the operation $(\deltabf\wedge\Fb) \rintprod \Fb $ and write it out in terms of components, i.~e., 
\begin{align}
(\deltabf\wedge\Fb) \rintprod \Fb  &= 
\Biggl( \sum_{\substack{j,B\\j\not\in B}} \Delta_{j,j} \partial_j F_B \sigma(j,B) \ebf_{j+B} \Biggr) \rintprod \Biggl(\sum_A F_A \ebf_A \Biggr) \\
&= \sum_{\substack{j,A,B\\A\in j+B}} \Delta_{j,j} F_A \partial_j F_B \sigma(j,B) \Delta_{A,A} \sigma(A, j+B\setminus A) \, \ebf_{j+B\setminus A}.
\label{eq:A21}
\end{align}
Again noting that $j+B\setminus A = i$ and that $A+i = B+j$, from~\eqref{eq:A21} we obtain
\begin{align}
(\deltabf\wedge\Fb) \rintprod \Fb  &= 
\sum_{i,j} \sum_{\substack{A,B\\A+ i = B+j}} F_A \partial _j F_B \, \sigma(i,A) \sigma(j,B)(-1)^{r} \Delta_{A,A}\Delta_{j,j} \, \ebf _i ,
\end{align}
proving the equivalence with \eqref{eq:deltaT2} and therefore proving~\eqref{eq:delta-T2}. 

Combining~\eqref{eq:delta-T1} and~\eqref{eq:delta-T2} and defining the stress-energy-momentum tensor $\Tb$ of the Maxwell field $\mf$ as $\Tb = -(\Tb_\odot  + \Tb_\owedge)$, we find that the interior derivative of the tensor $\Tb$ 
satisfies the following formula
\begin{equation}\label{eq:full-Lorentz}
\deltabf\lintprod\Tb +  (\deltabf\lintprod \mf)\lintprod \mf + (\deltabf\wedge \mf)\rintprod\mf = 0.
\end{equation}
Therefore, from the Maxwell equations, that is $\deltabf\wedge\mf = 0$ and $\deltabf\lintprod\mf = \sd$, we recover the conservation law for energy-momentum relating the Lorentz force $\fb$~\eqref{eq:lorentz} and the stress-energy-momentum tensor, 
\begin{equation}
	\fb + \deltabf\lintprod \Tb = 0.
\end{equation}

\subsection{Explicit Formulas for the Tensor Components for Generic $r$}
\label{app:components}

Starting with~\eqref{eq:Tij-int}, we note that 
\begin{align}
	\ebf_{i}\lintprod \mf &= \sum_{I\in\mathcal{I}_{r}}\Delta_{ii}\sigma(I\setminus i,i)\mfi_I \ebf_{I\setminus i}\\
	\mf\rintprod \ebf_{j} &= \sum_{J\in\mathcal{I}_{r}}\Delta_{jj}\sigma(j,J\setminus j)\mfi_J \ebf_{J\setminus j},
\end{align}
for any pair of $i$ and $j$, and therefore 
\begin{align}
 \mf\odot\mf\bigr|_{ij} &= \frac{1}{2}\sum_{I,J\in\mathcal{I}_{r}}\sigma(I\setminus i,i)\sigma(j,J\setminus j)\mfi_I \mfi_J \ebf_{I\setminus i}\cdot\ebf_{J\setminus j} \\
 &= \frac{1}{2}\sum_{L\in\mathcal{I}_{r-1}:i,j\not\in L}\sigma(L,i)\sigma(j,L)\mfi_{i+L} \mfi_{j+L} \Delta_{LL},\label{eq:A44}
\end{align} 
where we have defined the set $L$ such that $I\setminus i = J\setminus j = L$. The summation contains $\smash{\binom{k+n-2}{r-1}}$ non-zero terms. For $i = j$, and using that $\sigma(L,i)\sigma(i,L) = (-1)^{r-1}$, it can be evaluated as
\begin{align}
 \mf\odot\mf\bigr|_{ii} &= (-1)^{r-1}\frac{1}{2}\sum_{L\in\mathcal{I}_{r-1}:i\notin L}\mfi_{i+L}^2\Delta_{LL},
\end{align} 

Moving on to~\eqref{eq:Tij-ext}, we note that
\begin{align}
	\ebf_{i}\wedge \mf &= \sum_{I\in\mathcal{I}_{r}} \mfi_I\sigma(i,I)\ebf_{i+I} \\
	\mf\wedge \ebf_{j} &= \sum_{J\in\mathcal{I}_{r}} \mfi_J\sigma(J,j)\ebf_{j+J},
\end{align}
and we study the cases such that the lists $i+I$ and $j+J$ coincide. First, if $i = j$, we have to sum over $I = J$ such that $i \notin I$, that is 
\begin{align}
	\mf\owedge\mf\bigr|_{ii} &= \frac{1}{2}\sum_{I\in\mathcal{I}_{r}: i \notin I}\mfi_I^2\sigma(i,I)\sigma(I,i)\Delta_{ii}\Delta_{II} \\
	&= (-1)^{r}\frac{1}{2}\sum_{I\in\mathcal{I}_{r}: i \notin I}\mfi_I^2\Delta_{ii}\Delta_{II},
\end{align}
because $\sigma(i,I)\sigma(I,i) = (-1)^r$. Second, if $i\neq j$, we can find an $L \in\mathcal{I}_{r-1}$ such that $L = I\setminus j = J  \setminus i$ and $i,j\neq L$. Then,
\begin{align}
	 \mf\owedge\mf\bigr|_{ij} &= \frac{1}{2}\Delta_{ii}\Delta_{jj}\sum_{L \in\mathcal{I}_{r-1}} \mfi_{L+j}\mfi_{L+i}\sigma(i,L+j)\sigma(L+i,j)\ebf_{j+L+i}\cdot\ebf_{j+L+i} \\
	&= \frac{1}{2}\sum_{L \in\mathcal{I}_{r-1}} \mfi_{L+j}\mfi_{L+i}\sigma(i,L+j)\sigma(L+i,j)\Delta_{LL}.\label{eq:A51}
\end{align}
The product $\sigma(i,L+j)\sigma(L+i,j)$ is equal to $\sigma(L,i)\sigma(j,L)$. To prove it, we write the relation $\sigma(i,L+j)\sigma(L+i,j) = \sigma(L,i)\sigma(j,L)$ and we first multiply both sides by $\sigma(L+i,j)\sigma(j,L)$ to obtain $\sigma(i,L+j)\sigma(j,L) = \sigma(L+i,j)\sigma(L,i)$, namely the permutations sorting the lists $(i,j,L)$ and $(L,i,j)$ respectively. Secondly, we note that $\sigma(i,j,L) = (-1)^{2(r-1)}\sigma(L,i,j) = \sigma(L,i,j)$.

Combining~\eqref{eq:A44} and~\eqref{eq:A51} into $\Tb\bigr|_{ij} = -\mf\odot\mf\bigr|_{ij}-\mf\owedge\mf\bigr|_{ij}$, we have
\begin{align}
	T_{ii} &= \frac{(-1)^{r}}{2}\Delta_{ii}\Biggl(\sum_{I\in\mathcal{I}_{r}: i \in I}\mfi_I^2\Delta_{II} - \sum_{I\in\mathcal{I}_{r}: i \notin I}\mfi_{I}^2\Delta_{II}\Biggr) \\
	T_{ij} &= -\sum_{L\in\mathcal{I}_{r-1}}\sigma(L,i)\sigma(j,L)\mfi_{i+L} \mfi_{j+L} \Delta_{LL}. 
\end{align}

%
%
%

\subsection{Stokes Theorem for the Interior Derivative of a Bitensor}
\label{app:stokes_bitensor}

Considering a bitensor field
\begin{equation}
\Tb = \sum_i T_{ii}\ubf_{ii} + \sum_{i< j}T_{ij}\ubf_{ij} ,
\end{equation}
then the Stokes theorem we wish to prove states that
\begin{equation}
 \int_{\partial\mathcal{V}^{n+k}}  \drm ^{n+k-1} \xb^\hodgeinv \lintprod \Tb
= \int_{\mathcal{V}^{n+k}} \left( \drm^{n+k} x \right)^\hodgeinv \lintprod \left(\partial \lintprod \Tb  \right) .
\end{equation}
The proof will follow the reasoning operating for the vector field \cite[Sect.~3.5]{colombaro2019introductionSpaceTimeExteriorCalculus}, so we start expanding the integrand on the right-hand side
\begin{align}
\drm ^{n+k-1} \xb^\hodgeinv \lintprod \Tb &= 
\left(\sum_{I_{n+k-1} : q\not\in I} \drm x_I \Delta_{qq} \sigma(q,I) \ebf_q \right) \lintprod\left( \sum_i T_{ii}\ubf_{ii} + \sum_{i< j}T_{ij}\ubf_{ij} \right)	\\
&= \sum_{i,I_{n+k-1} : i\not\in I} T_{ii} \drm x_I \sigma(i,I) \, \ebf_i
+\sum_{i<j}\sum_{I_{n+k-1} : j\not\in I} T_{ij} \drm x_I \sigma(j,I) \, \ebf_i \notag \\&\qquad + \sum_{i<j}\sum_{I_{n+k-1} : i\not\in I} T_{ij} \drm x_I \sigma(i,I) \, \ebf_j .
\end{align}
The last term can be rewritten by changing the indices $i\longleftrightarrow j$ and using the property of symmetry $T_{ji}=T_{ij}$ as
\begin{align}
\sum_{i>j}\sum_{I_{n+k-1} : j\not\in I} T_{ij} \drm x_I \sigma(j,I) \, \ebf_i ,
\end{align}
and finally, the three terms can be compacted in
\begin{equation}
\drm ^{n+k-1} \xb^\hodgeinv \lintprod \Tb = \sum _{i,j} \sum_{I_{n+k-1} : j\not\in I} T_{ij} \drm x_I \sigma(j,I) \, \ebf_i .
\end{equation}
Then, taking the exterior derivative \cite[Sect.~36.B]{arnold1989mathematicalMethods}, we get
\begin{align}
\drm\left(\drm ^{n+k-1} \xb^\hodgeinv \lintprod \Tb\right) &= \sum _{i,j} \sum_{I_{n+k-1} : j\not\in I} \partial_j T_{ij} \drm x_{\varepsilon(j, I)} \sigma^2(j,I) \, \ebf_i 
\\
&= 
\sum _{i,j} \sum_{L_{n+k}} \partial_j T_{ij} \drm x_{L} \, \ebf_i .\label{eq:A64} 
\end{align}

On the other hand, regarding the right-hand side, we first evaluate
\begin{align}
\partial \lintprod \Tb &= \left( \sum_h \Delta_{hh} \partial_h \ebf_h \right) \lintprod \left( \sum_i T_{ii}\ubf_{ii} + \sum_{i< j}T_{ij}\ubf_{ij} \right) \\
&= \sum_i \partial_i T_{ii} \ebf_i + \sum_{i<j} \partial_j T_{ij} \ebf_i + \sum_{i<j} \partial_i T_{ij} \ebf_j 	\\
&= \sum_{i,j} \partial_j T_{ij} \ebf_i
\end{align}
and the differential
\begin{align}
\left( \drm^{n+k} \xb \right)^\hodgeinv = \left( \sum_{L_{n+k}} \drm x_L \ebf_L \right)^\hodgeinv = \sum _{L_{n+k}} \drm x_L.
\end{align}
Then, we get
\begin{equation}
\left( \drm^{n+k} \xb \right)^\hodgeinv \lintprod \left(\partial \lintprod \Tb  \right) = \sum _{L_{n+k}} \sum_{i,j} \partial_j T_{ij}\drm x_L \, \ebf_i,
\end{equation}
namely~\eqref{eq:A64} and thereby proving the stated Stokes' Theorem.

\subsection{Flux of the Stress-Energy-Momentum Tensor}
\label{app:flux-T}

The flux of the field~\eqref{eq:T_Tij} across the boundary $\partial \mathcal{V}_\ell^{k+n}$, denoted by $\Phi_{\partial \mathcal{V}_\ell^{k+n}} (\Tb)$, is given by the integral in~\eqref{eq:flux_Til},
\begin{equation}
\Phi_{\partial \mathcal{V}_\ell^{k+n}} (\Tb) = \int_{-\infty}^{+\infty} \dots \int_{-\infty}^{+\infty} \drm x_{\ell^c} \Delta_{\ell\ell}\sigma (\ell , \ell^c) \, \ebf_\ell \lintprod \Tb.
\label{eq:proofT3}
\end{equation}
The r.h.s.~of \eqref{eq:proofT3} is computed w.r.t.~$x_{\ell^c}$, being $\ell^c$ the set of indices excluding $\ell$.

We next write the flux~\eqref{eq:proofT3} in terms of the Fourier transform of $\Fb$, denoted as $\hat{\mf} (\xibf)$ as in~\eqref{eq:F-fourier}. Assuming that the Fourier transform of $\Fb$ is supported only in the set $\xibf\cdot\xibf=0$, as postulated in~\eqref{eq:fourtransF}, we express $\Fb$ as
\begin{equation}
\mf (\xb) = \int_{-\infty}^{+\infty}\dots\int_{-\infty}^{+\infty}
\drm^{k+n}\xibf \, \delta( \xibf\cdot\xibf ) \,
e^{j2\pi\xibf \cdot \xb} \, \hat{\mf} (\xibf).
\label{eq:proofT4}
\end{equation}
Inserting~\eqref{eq:proofT4} in~\eqref{eq:totalT} and using the linearity properties of $\odot$ and $\owedge$, we obtain that the stress-energy-momentum tensor $\Tb$ can be written as
\begin{equation}
\Tb = -\frac{1}{2}
\int_{-\infty}^{+\infty}\dots\int_{-\infty}^{+\infty}
\drm^{k+n}\xibf \drm^{k+n}\xibf' \,
\delta(\xibf\cdot\xibf) \delta(\xibf'\cdot\xibf')  \,
e^{j2\pi(\xibf+ \xibf')\cdot \xb}
	\bigl( \hat{\mf}(\xibf)\odot \hat{\mf}(\xibf') + \hat{\mf}(\xibf)\owedge \hat{\mf}(\xibf') \bigr) .	
\label{eq:proofT5}	
\end{equation}
Using~\eqref{eq:proofT5} back in~\eqref{eq:proofT3}, we obtain that
\begin{multline}
\Phi_{\partial \mathcal{V}_\ell^{k+n}} (\Tb) 
=
- \frac{1}{2}\Delta_{\ell\ell}\sigma (\ell , \ell^c)
\int_{-\infty}^{+\infty} \dots \int_{-\infty}^{+\infty} 
\drm^{k+n} \xibf \drm^{k+n}\xibf' \drm x_{\ell^c}  
\\
\delta(\xibf\cdot\xibf) \delta(\xibf'\cdot\xibf')  \,
e^{j2\pi(\xibf+\xibf')\cdot \xb}
\ebf_\ell \lintprod \bigl( \hat{\mf}(\xibf)\odot \hat{\mf}(\xibf') + \hat{\mf}(\xibf)\owedge \hat{\mf}(\xibf') \bigr).	\label{eq:proofT6}
\end{multline}

Since the integration w.r.t.~$x_{\ell^c}$ only acts on the exponential term in~\eqref{eq:proofT6}, interchanging the integration order and using the definition of the delta function, we can find the inverse Fourier transform of the exponential as
\begin{equation}
\int_{-\infty}^{+\infty} \dots \int _{-\infty}^{+\infty}
\drm x_{\ell^c}  \,
e^{j2\pi(\xibf+\xibf')\cdot \xb}
= \prod_{m\ne \ell} \delta (\xi_m + \xi_m') 
e^{j2\pi(\xi_{\ell} + \xi'_{\ell} )x_{\ell} \Delta_{\ell\ell}}.
\label{eq:proofT8}
\end{equation}
Plugging the r.h.s. of~\eqref{eq:proofT8} in~\eqref{eq:proofT6} and defining $\varphi(\xibf,\xibf')$ as
\begin{equation}
	\varphi(\xibf,\xibf') = -\frac 12 \Delta_{\ell\ell}\sigma (\ell , \ell^c) e^{j2\pi(\xi_{\ell} + \xi'_{\ell} )x_{\ell} \Delta_{\ell\ell}} \ebf_\ell \lintprod \bigl( \hat{\mf}(\xibf)\odot \hat{\mf}(\xibf') + \hat{\mf}(\xibf)\owedge \hat{\mf}(\xibf') \bigr), 
	\label{eq:T_Tij4}
\end{equation} 
we write the flux as
\begin{equation}
\Phi_{\partial \mathcal{V}_\ell^{k+n}} (\Tb) = \int_{-\infty}^{+\infty} \dots \int _{-\infty}^{+\infty}
\drm^{k+n} \xibf \drm^{k+n} \xibf'\,
\delta(\xibf\cdot\xibf) \delta(\xibf'\cdot\xibf')  \prod_{m\ne \ell} \delta (\xi_m + \xi_m') 
 \varphi(\xibf, \xibf').
 \label{eq:proofT9}
\end{equation}

In order to solve the integration w.r.t.~$\xi_\ell$, we rewrite the condition  $\xibf \cdot\xibf=0$ as
\begin{equation}
\Delta_{\ell\ell}\xi_\ell^2 + \xibf_{\bar\ell}\cdot\xibf_{\bar\ell} = 0,
\label{eq:proofT7}
\end{equation} 
where $\xibf_{\bar\ell} = \xibf -\xi_\ell\ebf_\ell$.
We can solve this equation for $\xi_\ell$ as long as $-\Delta_{\ell\ell}  \xibf_{\bar\ell}\cdot\xibf_{\bar\ell} \geq 0$, in which case we define $\chi_\ell$ as the positive root of the equation
\begin{equation}
\chi_\ell^2 = -\Delta_{\ell\ell}  \xibf_{\bar\ell}\cdot\xibf_{\bar\ell},
\label{eq:T_Tij1}
\end{equation}
and we  thus take for $\xi_\ell$ the two possible values $\xi_\ell  = \pm\chi_\ell$.
%
We similarly have the analogous versions of~\eqref{eq:proofT7} and~\eqref{eq:T_Tij1} for $\xi_\ell'$. 

Using \cite[p.~184]{Gelfand1966} w.r.t.~the integration variables $\xi_\ell$ and $\xi_\ell'$ and the limitation in the integration range, equation~\eqref{eq:proofT9} is expressed as  
\begin{equation}
\Phi_{\partial \mathcal{V}_\ell^{k+n}} (\Tb) =
\int_{\substack{\Delta_{\ell\ell} \xibf_{\bar\ell} \cdot \xibf_{\bar\ell}\le 0\\\Delta_{\ell\ell} \xibf'_{\bar\ell} \cdot \xibf'_{\bar\ell}\le 0}}
\drm \xi_{\ell^c} \drm \xi'_{\ell^c}\, \frac{1}{4\chi_\ell\chi'_\ell}  \prod_{m\ne \ell} \delta (\xi_m + \xi_m') 
 \sum_{\xi_\ell= \pm \chi_\ell, \xi'_\ell = \pm \chi_\ell'}
\varphi(\xibf, \xibf').
\label{eq:T_Tij2}
\end{equation}
Since the remaining Dirac delta function conditions imply that $\xi'_m=-\xi_m$ for every $m\neq\ell$, we also have that $\xi'_\ell=\pm\chi_\ell$. To further deal with the four terms in the summation in~\eqref{eq:T_Tij2}, we define the  vectors
\begin{align}
\xibf_+ &=(\xi_0, \dots, \xi_{\ell-1}, \chi_\ell,  \xi_{\ell+1},\dots, \xi_{k+n-1}), \label{eq:xiplus} \\
\xibf_- &=(\xi_0, \dots, \xi_{\ell-1},-\chi_\ell, \xi_{\ell+1},\dots, \xi_{k+n-1}), \label{eq:ximinus}  
\end{align}
and the counterparts $\xibf'_+=-\xibf_-$ and  $\xibf'_-=-\xibf_+$. Using these definitions to solve the integration w.r.t.~$\xi'_{\ell^c}$, we obtain that
\begin{equation}
\Phi_{\partial \mathcal{V}_\ell^{k+n}} (\Tb) =  	\int_{\Delta_{\ell\ell} \xibf_{\bar\ell} \cdot \xibf_{\bar\ell}\le 0}
\drm \xi_{\ell^c} \, \frac{1}{4\chi_\ell^2} \bigl(
\varphi(\xibf_+,\xibf_+') +		
\varphi(\xibf_+,\xibf_-') +		
\varphi(\xibf_-,\xibf_+') +
\varphi(\xibf_-,\xibf_-') 
 \bigr).
 \label{eq:T_Tij3}
\end{equation}

It remains to study the four summands in~\eqref{eq:T_Tij3} by exploiting the properties of exterior algebra. We start by writing $\varphi(\xibf_+,\xibf_+')$ from its definition in~\eqref{eq:T_Tij4} and use the fact that $\hat{\mf}(\xibf'_+)=\hat{\mf}(-\xibf_-)=\hat{\mf}^*(\xibf_-)$ to obtain
\begin{equation}
	\varphi(\xibf_+,\xibf_-) = - \frac 12 \Delta_{\ell\ell}\sigma (\ell , \ell^c) e^{j4\pi\chi_{\ell}x_{\ell} \Delta_{\ell\ell}} \ebf_\ell \lintprod \tens_1
	\label{eq:integr-region0}
\end{equation}
where for the sake of clarity we defined the tensor $\tens_1$ as
\begin{equation}
	\tens_1 =  \hat{\mf}(\xibf_+) \odot \hat{\mf}^* (\xibf_-) + \hat{\mf}(\xibf_+) \owedge \hat{\mf}^* (\xibf_-).
\label{eq:tens1}
\end{equation}
Using the definitions of $\odot$ and $\owedge$ in~\eqref{eq:Tij-int} and~\eqref{eq:Tij-ext} respectively, and the identity~\eqref{eq:text-equiv-wedge-int}, we may write the $ij$-th component of $\tens_1$ as
\begin{align}
\tensc_{1,ij} &=  \Delta_{ii}\Delta_{jj} \bigl( (\ebf_{i}\lintprod \hat{\mf}(\xibf_+))\cdot (\hat{\mf}^*(\xibf_-)\rintprod \ebf_{j}) + (\ebf_{i}\wedge\hat{\mf}(\xibf_+))\cdot(\hat{\mf}^*(\xibf_-) \wedge \ebf_{j}) \bigr) \\
&=
 \Delta_{ii}\Delta_{jj} 
 \bigl(
 \underbrace{
(\ebf_{i}\lintprod \hat{\mf}(\xibf_+))\cdot (\hat{\mf}^*(\xibf_-)\rintprod \ebf_{j})
	}_{\alpha_{1,ij}} 
  + 
   \underbrace{
(-1)^r\Delta_{ij}   \hat{\mf}(\xibf_+) \cdot  \hat{\mf}^*(\xibf_-) 
	}_{\beta_{1,ij}} 
\notag	\\	
&\qquad+
 \underbrace{
(\ebf_{j}\lintprod\hat{\mf}(\xibf_+))\cdot(\hat{\mf}^*(\xibf_-) \rintprod \ebf_{i})
	}_{\alpha_{1,ji}} 
 \bigr).
 \label{eq:fromI1to.}
\end{align}

It will prove convenient to study equation~\eqref{eq:fromI1to.} in terms of the $(r-1)$-vector potential $\Ab$, which is related to $\Fb$ as
in \eqref{eq:field-potential}, or in the Fourier domain,
\begin{equation}
\hat{\mf} (\xibf) = 2\pi j \, \xibf \wedge \hat{\Ab} (\xibf),
\label{eq:T_Tij6}
\end{equation}
where $\hat{\Ab} (\xibf)$ denotes the Fourier transform of $\Ab$. Substituting~\eqref{eq:T_Tij6} in the definitions of $\alpha_{1,ij}$ and $\beta_{1,ij}$ in~\eqref{eq:fromI1to.} and using the identity~\eqref{eq:comm-intprod}, we obtain that
\begin{gather}
	\alpha_{1,ij} = 
4\pi^2
(-1)^{r-1} \bigl(\ebf_{i}\lintprod (\xibf_+ \wedge \hat{\Ab}(\xibf_+))\bigr)\cdot \bigl( \ebf_{j} \lintprod ( \xibf_- \wedge\hat{\Ab}^*(\xibf_-))\bigr)
		 \label{eq:proof-I1-l1}\\
\beta_{1,ij} = 
4\pi^2 (-1)^r\Delta_{ij} (\xibf_+ \wedge \hat{\Ab}(\xibf_+)) \cdot  (\xibf_- \wedge \hat{\Ab}^*(\xibf_-)) 
.	\label{eq:proof-I1-l3}
\end{gather}


We start by expanding $\alpha_{1,ij}$. Using the identiy~\eqref{eq:equiv-lint-wedge}, we get 
\begin{multline}
	\alpha_{1,ij} = 4\pi^2
(-1)^{r-1} \Bigl((-1)^{r-1}(\ebf_i \cdot \xibf_+) \hat{\Ab}(\xibf_+) + \xibf_+ \wedge \bigl(\ebf_i \lintprod\hat{\Ab}(\xibf_+) \bigr)\Bigr)
 \\
\cdot
\Bigl((-1)^{r-1}(\ebf_j \cdot \xibf_-) \hat{\Ab}^*(\xibf_-) + \xibf_- \wedge (\ebf_j \lintprod\hat{\Ab}^*(\xibf_-) ) \Bigr).
\end{multline}
Computing the products, rearranging terms and using the relations~\eqref{eq:text-equiv-wedge-int}, \eqref{eq:text-equiv-double-interior}--\eqref{eq:text-equiv-double-interior2} and~\eqref{eq:text-wedge-dot-} in various places, we obtain
\begin{align}
\alpha_{1,ij} =4\pi^2
\Bigl( &
(-1)^{r-1}\Delta_{ii}\Delta_{jj}\xi_{+,i}\xi_{-,j} \hat{\Ab}(\xibf_+)\cdot \hat{\Ab}^*(\xibf_-) 
\notag
\\
&+ \Delta_{ii}\xi_{+,i}
\bigl(\ebf_j \lintprod \hat{\Ab}^*(\xibf_-) \bigr) \cdot \bigl(\hat{\Ab}(\xibf_+) \rintprod \xibf_- \bigr)	\notag
\\
& +\Delta_{jj}\xi_{-,j}
\bigl( \ebf_i \lintprod \hat{\Ab}(\xibf_+)\bigr) \cdot \bigl(\hat{\Ab}^*(\xibf_-) \rintprod \xibf_+  \bigr)
\notag
\\
&+(-1)^{r-1}(\xibf_+\cdot\xibf_-) \bigl(\ebf_i \lintprod \hat{\Ab}(\xibf_+)\bigr)\cdot\bigl(\ebf_j \lintprod \hat{\Ab}^*(\xibf_-)\bigr)
\notag
\\
&
+\bigl( \ebf_i \lintprod \bigl(\xibf_- \lintprod \hat{\Ab}(\xibf_+)\bigr) \bigr)\cdot
\bigl(\ebf_j \lintprod \bigl(\hat{\Ab}^*(\xibf_-)\rintprod\xibf_+\bigr) \bigr)
\Bigr).
\label{eq:T_Tij9}
\end{align}

We next simplify the terms of the form $\xibf_- \lintprod \hat{\Ab}(\xibf_+)$ and $\hat{\Ab}^*(\xibf_-)\rintprod\xibf_+$. To do so, we note that $\xibf_+$ and $\xibf_-$ respectively given in~\eqref{eq:xiplus} and~\eqref{eq:ximinus} are related as
\begin{equation}
\xibf_+ =  \xibf_- + 2\chi_\ell\ebf_\ell.
\label{eq:T_Tij7}
\end{equation}
Recalling that the gauge condition in the Fourier domain is given by
\begin{equation}
\xibf_+ \lintprod \hat{\Ab}(\xibf_+)= 0 = \xibf_- \lintprod \hat{\Ab}^*(\xibf_-),
\label{eq:T_Tij8}
\end{equation}
equations~\eqref{eq:T_Tij7} and~\eqref{eq:T_Tij8} imply that
\begin{gather}
\xibf_+ \lintprod \hat{\Ab}^*(\xibf_-) = \xibf_- \lintprod \hat{\Ab}^*(\xibf_-) + 2\chi_\ell\ebf_\ell \lintprod \hat{\Ab}^*(\xibf_-)
= 2\chi_\ell\ebf_\ell \lintprod \hat{\Ab}^*(\xibf_-) , \label{eq:xi+Astar-}
\\
\xibf_- \lintprod \hat{\Ab}(\xibf_+) = \xibf_+ \lintprod \hat{\Ab}(\xibf_+) - 2\chi_\ell\ebf_\ell \lintprod \hat{\Ab}(\xibf_+)
= -2\chi_\ell\ebf_\ell \lintprod \hat{\Ab}(\xibf_+). \label{eq:xi-A+}
\end{gather}
Similar relations are obtained for the right interior product. Applying~\eqref{eq:xi+Astar-} and \eqref{eq:xi-A+} into~\eqref{eq:T_Tij9}, we obtain for $\alpha_{1,ij}$ that
\begin{align}
\alpha_{1,ij} =4\pi^2
\Bigl( &
(-1)^{r-1}\Delta_{ii}\Delta_{jj}\xi_{+,i}\xi_{-,j} \hat{\Ab}(\xibf_+)\cdot \hat{\Ab}^*(\xibf_-) 
\notag	\\
&- 2\Delta_{ii}\xi_{+,i} \chi_\ell
\bigl(\ebf_j \lintprod \hat{\Ab}^*(\xibf_-) \bigr) \cdot \bigl(\hat{\Ab}(\xibf_+) \rintprod \ebf_\ell \bigr)
\notag
\\
&+2\Delta_{jj}\xi_{-,j} \chi_\ell
\bigl( \ebf_i \lintprod \hat{\Ab}(\xibf_+)\bigr) \cdot \bigl(\hat{\Ab}^*(\xibf_-) \rintprod \ebf_\ell  \bigr)
\notag	\\
&+(-1)^{r-1}\bigl(-2\Delta_{\ell\ell}\chi_\ell^2\bigr) \bigl(\ebf_i \lintprod \hat{\Ab}(\xibf_+)\bigr)\cdot\bigl(\ebf_j \lintprod \hat{\Ab}^*(\xibf_-)\bigr)
\notag
\\
&
-4 \chi_\ell^2 \bigl( \ebf_i \lintprod \bigl(\ebf_\ell \lintprod \hat{\Ab}(\xibf_+)\bigr) \bigr)\cdot
\bigl(\ebf_j \lintprod \bigl(\hat{\Ab}^*(\xibf_-)\rintprod\ebf_\ell\bigr) \bigr)		
\Bigr).
\label{eq:alpha}
\end{align}

For $\beta_{1,ij}$, we first use~\eqref{eq:text-equiv-wedge-int} directly into \eqref{eq:proof-I1-l3} so that it is written as
\begin{align}
\beta_{1,ij} &= 4\pi^2
(-1)^r\Delta_{ij}(-1)^{r-1}
\bigl( (-1)^{r-1} (\xibf_+ \cdot \xibf_-) (\hat{\Ab}(\xibf_+)\cdot \hat{\Ab}^*(\xibf_-) ) 
\notag		\\
&\qquad+ 
(\xibf_+\lintprod \hat{\Ab}^*(\xibf_-))\cdot(\hat{\Ab}(\xibf_+)\rintprod \xibf_-)
\bigr)
\\
&= 4\pi^2 (-1)^r\Delta_{ij} (\xibf_+ \cdot \xibf_-) \bigl(\hat{\Ab}(\xibf_+)\cdot \hat{\Ab}^*(\xibf_-) \bigr)
\notag		\\
&\qquad- 4\pi^2\Delta_{ij} \bigl(\xibf_+\lintprod \hat{\Ab}^*(\xibf_-)\bigr)\cdot\bigl(\hat{\Ab}(\xibf_+)\rintprod \xibf_-\bigr).
\end{align}
In view of~\eqref{eq:T_Tij7} and $\xibf_+ \cdot\xibf_+ = 0$, the $\xibf_+\cdot\xibf_-$ term in the previous equation equals
\begin{equation} \label{eq:xibf+and-}
\xibf_+\cdot\xibf_- = \xibf_+\cdot\xibf_+ - 2\chi_\ell\,\xibf_+\cdot\ebf_\ell = -2 \Delta_{\ell\ell} \chi^2_\ell.
\end{equation}
Therefore, using~\eqref{eq:xi+Astar-},~\eqref{eq:xi-A+} and~\eqref{eq:xibf+and-} we finally obtain
\begin{equation}
\beta_{1,ij} = 8\pi^2 \chi_\ell^2 \Delta_{ij} \Bigl((-1)^{r-1} \Delta_{\ell\ell} \hat{\Ab}(\xibf_+)\cdot \hat{\Ab}^*(\xibf_-)
+2 \bigl(\ebf_\ell\lintprod \hat{\Ab}^*(\xibf_-)\bigr)\cdot\bigl(\hat{\Ab}(\xibf_+)\rintprod \ebf_\ell\bigr)\Bigr).
\label{eq:gamma1}
\end{equation}

Although we derived expressions of $\alpha_{1,ij}$, $\alpha_{1,ji}$ and $\beta_{1,ij}$ in~\eqref{eq:alpha} and~\eqref{eq:gamma1}, needed to obtain $\tensc_{1,ij}$ in~\eqref{eq:fromI1to.} for arbitrary $ij$, we are only interested in such terms containing the $\ell$-th component, since $\varphi(\xibf_+,\xibf_+')$ in~\eqref{eq:integr-region0} involves computing the quantity
\begin{align}
\ebf_\ell \lintprod \tens_1 
&= \ebf_\ell \lintprod \sum_{i\le j} \tensc_{1,ij} \ubf_{ij}
= \tensc_{1,\ell\ell} \Delta_{\ell\ell} \ebf_\ell
+ \sum_{i\ne \ell} (\tensc_{1,i\ell} + \tensc_{1,\ell i}) \Delta_{\ell\ell} \ebf_i.
\label{eq:integr-region3}
\end{align}


We start with the first case in which $i=\ell\ne j$. Using that $\ebf_\ell\lintprod\ebf_\ell=0$, from~\eqref{eq:alpha} we get
\begin{align}
\alpha_{1,\ell j} =4\pi^2
\Bigl( &
(-1)^{r-1}\Delta_{\ell\ell}\Delta_{jj} \xi_{+,\ell}\xi_{-,j} \hat{\Ab}(\xibf_+)\cdot \hat{\Ab}^*(\xibf_-)
\notag
\\ 
&-2 \Delta_{\ell\ell}\xi_{+,\ell} \chi_\ell
\bigl(\ebf_j \lintprod \hat{\Ab}^*(\xibf_-) \bigr) \cdot \bigl( \hat{\Ab}(\xibf_+) \rintprod \ebf_\ell \bigr) \notag
\\
&+2\Delta_{jj}\xi_{-,j} \chi_\ell
\bigl( \ebf_\ell \lintprod \hat{\Ab}(\xibf_+)\bigr) \cdot \bigl(\hat{\Ab}^*(\xibf_-) \rintprod \ebf_\ell  \bigr)
\notag
\\
&+2(-1)^{r}\Delta_{\ell\ell}\chi^2_\ell
\bigl(\ebf_\ell \lintprod \hat{\Ab}(\xibf_+)\bigr)\cdot\bigl(\ebf_j \lintprod \hat{\Ab}^*(\xibf_-)\bigr) 
\Bigr).\label{eq:alpha1b}
\end{align}
Similarly, we also have
\begin{align}
\alpha_{1, j\ell} =4\pi^2
\Bigl( &
(-1)^{r-1}\Delta_{\ell\ell}\Delta_{jj} \xi_{+,j}\xi_{-,\ell} \hat{\Ab}(\xibf_+)\cdot \hat{\Ab}^*(\xibf_-) 
\notag
\\
&-2 \Delta_{jj}\xi_{+,j} \chi_\ell
\bigl(\ebf_\ell \lintprod \hat{\Ab}^*(\xibf_-) \bigr) \cdot \bigl( \hat{\Ab}(\xibf_+) \rintprod \ebf_\ell \bigr) \notag
\\
&+2\Delta_{\ell\ell}\xi_{-,\ell} \chi_\ell
\bigl( \ebf_j \lintprod \hat{\Ab}(\xibf_+)\bigr) \cdot \bigl(\hat{\Ab}^*(\xibf_-) \rintprod \ebf_\ell  \bigr)
\notag
\\
&+2(-1)^{r}\Delta_{\ell\ell}\chi^2_\ell
\bigl(\ebf_j \lintprod \hat{\Ab}(\xibf_+)\bigr)\cdot\bigl(\ebf_\ell \lintprod \hat{\Ab}^*(\xibf_-)\bigr) 
\Bigr).\label{eq:beta1b}
\end{align}
Furthermore, the fact that $\Delta_{ij}=0$ for $i\neq j$ implies from~\eqref{eq:gamma1} that
\begin{align}
\beta_{1,\ell j} = 0.\label{eq:gamma1b}
\end{align}
Combining~\eqref{eq:alpha1b},~\eqref{eq:beta1b} and~\eqref{eq:gamma1b} in the initial expression of $\tensc_{1,\ell j}$ in~\eqref{eq:fromI1to.}, using that $\xi_{\pm,\ell} = \pm \chi_\ell$, and writing the right interior products as left interior products, we get
\begin{align}
\tensc_{1,\ell j} 
 &= 4\pi^2 \Delta_{\ell\ell}\Delta_{jj} 
\Bigl(
(-1)^{r-1}\Delta_{\ell\ell}\Delta_{jj}\hat{\Ab}(\xibf_+)\cdot \hat{\Ab}^*(\xibf_-) \chi_\ell \left( \xi_{-,j} - \xi_{+,j} \right)
\notag
\\
&\qquad+2(-1)^r \Delta_{jj}  
\bigl( \ebf_\ell \lintprod \hat{\Ab}(\xibf_+)\bigr) \cdot \bigl( \ebf_\ell \lintprod \hat{\Ab}^*(\xibf_-) \bigr)
\chi_\ell (\xi_{-,j}-\xi_{+,j})
\notag
\\
&\qquad+2 (-1)^r \Delta_{\ell\ell}
\bigl( \ebf_j \lintprod \hat{\Ab}(\xibf_+)\bigr) \cdot \bigl(\ebf_\ell \lintprod \hat{\Ab}^*(\xibf_-)   \bigr)(\chi^2_\ell-\chi^2_\ell)
\notag
\\
&\qquad+2(-1)^{r}\Delta_{\ell\ell}
\bigl(\ebf_\ell \lintprod \hat{\Ab}(\xibf_+)\bigr)\cdot\bigl(\ebf_j \lintprod \hat{\Ab}^*(\xibf_-)\bigr) (\chi^2_\ell-\chi^2_\ell)\Bigr).\label{proof-I1'-line1}
\end{align}
We note that the last two summands in the former equation trivially cancel out, whereas the remaining two also do so because $\xi_{-,j}=\xi_{+,j}$ for $j\neq\ell$. Hence,
\begin{equation}
\tensc_{1,\ell j} = 0 , \qquad j\ne \ell .
\end{equation}


We continue with the second case $j=\ell\ne i$ and we have, as in the first case
\begin{align}
\alpha_{1,i\ell} =4\pi^2
\Bigl( &
(-1)^{r-1}\Delta_{ii}\Delta_{\ell\ell} \xi_{+,i}\xi_{-,\ell} \hat{\Ab}(\xibf_+)\cdot \hat{\Ab}^*(\xibf_-) 
\notag	\\
&-2 (-1)^r\Delta_{ii}\xi_{+,i} \chi_\ell
\bigl(\ebf_\ell \lintprod \hat{\Ab}^*(\xibf_-) \bigr) \cdot \bigl( \ebf_\ell \lintprod \hat{\Ab}(\xibf_+) \bigr) \notag
\\
&+2(-1)^r\Delta_{\ell\ell}\xi_{-,\ell} \chi_\ell
\bigl( \ebf_i \lintprod \hat{\Ab}(\xibf_+)\bigr) \cdot \bigl(\ebf_\ell \lintprod \hat{\Ab}^*(\xibf_-) \bigr)
\notag	\\
&+2(-1)^{r}\Delta_{\ell\ell}\chi^2_\ell
\bigl(\ebf_i \lintprod \hat{\Ab}(\xibf_+)\bigr)\cdot\bigl(\ebf_\ell \lintprod \hat{\Ab}^*(\xibf_-)\bigr) 
\Bigr) , \label{eq:alpha1c}
\end{align}
\begin{align}
\alpha_{1,\ell i} =4\pi^2
\Bigl( &
(-1)^{r-1}\Delta_{ii}\Delta_{\ell\ell} \xi_{+,\ell}\xi_{-,i} \hat{\Ab}(\xibf_+)\cdot \hat{\Ab}^*(\xibf_-) 
\notag	\\
&-2(-1)^r \Delta_{\ell\ell}\xi_{+,\ell} \chi_\ell
\bigl(\ebf_\ell \lintprod \hat{\Ab}^*(\xibf_-) \bigr) \cdot \bigl( \hat{\Ab}(\xibf_+) \rintprod \ebf_\ell \bigr) \notag
\\
&+2(-1)^r \Delta_{ii}\xi_{-,i} \chi_\ell
\bigl( \ebf_\ell \lintprod \hat{\Ab}(\xibf_+)\bigr) \cdot \bigl(\ebf_\ell \lintprod \hat{\Ab}^*(\xibf_-) \bigr)
\notag	\\
&+2(-1)^{r}\Delta_{\ell\ell}\chi^2_\ell
\bigl(\ebf_\ell \lintprod \hat{\Ab}(\xibf_+)\bigr)\cdot\bigl(\ebf_i \lintprod \hat{\Ab}^*(\xibf_-)\bigr) 
\Bigr) , \label{eq:beta1c}
\end{align}
and
\begin{equation}
\beta_{1,i\ell} = 0.\label{eq:gamma1c}
\end{equation}

We rearrange~\eqref{eq:alpha1c},~\eqref{eq:beta1c} and~\eqref{eq:gamma1c} in the initial expression~\eqref{eq:fromI1to.}, using $\xi_{\pm,\ell = \pm \chi_\ell}$ and that $\xi_{-,i}=\xi_{+,i}$ for $i \neq \ell$, we obtain
It results to be zero since for $i \neq \ell$ we have $\xi_{-,i}=\xi_{+,i}$, namely
\begin{equation}
\tensc_{1,i\ell} = 0 , \qquad i\ne \ell .
\end{equation}

Regarding the last case $i=j=\ell$ we evaluate $\alpha_{1,\ell\ell} $ from~\eqref{eq:alpha} writing all the right interior products as left interior products
\begin{align}
\alpha_{1,\ell\ell} =4\pi^2
\Bigl( &
(-1)^{r-1} \xi_{+,\ell}\xi_{-,\ell} \hat{\Ab}(\xibf_+)\cdot \hat{\Ab}^*(\xibf_-) 
\notag	\\
&-2 (-1)^r\Delta_{\ell\ell}\xi_{+,\ell} \chi_\ell
\bigl(\ebf_\ell \lintprod \hat{\Ab}^*(\xibf_-) \bigr) \cdot \bigl( \ebf_\ell \lintprod \hat{\Ab}(\xibf_+) \bigr) \notag
\\
&+2(-1)^r\Delta_{\ell\ell}\xi_{-,\ell} \chi_\ell
\bigl( \ebf_\ell \lintprod \hat{\Ab}(\xibf_+)\bigr) \cdot \bigl(\ebf_\ell \lintprod \hat{\Ab}^*(\xibf_-) \bigr)
\notag	\\
&+2(-1)^{r}\Delta_{\ell\ell}\chi^2_\ell
\bigl(\ebf_\ell \lintprod \hat{\Ab}(\xibf_+)\bigr)\cdot\bigl(\ebf_\ell \lintprod \hat{\Ab}^*(\xibf_-)\bigr) 
\Bigr) ,
\label{eq:integr-region1}
\end{align}
and from~\eqref{eq:gamma1}
\begin{equation}
\beta_{1,\ell\ell} = 8\pi^2(-1)^r \chi_\ell^2\Bigl(-
  \hat{\Ab}(\xibf_+)\cdot \hat{\Ab}^*(\xibf_-)
+2 \Delta_{\ell\ell} \bigl(\ebf_\ell\lintprod \hat{\Ab}^*(\xibf_-)\bigr)\cdot\bigl( \ebf_\ell \lintprod\hat{\Ab}(\xibf_+) \bigr)	\Bigr).
\label{eq:integr-region2}
\end{equation}
We substitute $\alpha_{1,\ell\ell}=\beta_{1,\ell\ell}$ from~\eqref{eq:integr-region1} and $\gamma_{1,\ell\ell}$ from~\eqref{eq:integr-region1} into $T_{1,\ell\ell}$~\eqref{eq:fromI1to.} and considering that $\xi_{\pm_\ell} = \pm \chi_\ell$, we directly get
\begin{align}
\tensc_{1,\ell\ell} 
=  0 .
\end{align}
In conclusion, from~\eqref{eq:integr-region3} we realize that
\begin{equation}
\ebf_\ell \lintprod \tens_1 = 0 .
\label{eq:proofT30}
\end{equation}


We continue studying the second summand in~\eqref{eq:T_Tij3}. 
We consider $\varphi(\xibf_+,\xibf_-')$ using its definition in~\eqref{eq:T_Tij4} joint with the fact that
$\hat{\mf}(\xibf'_-) = \hat{\mf}(-\xibf_+) =\hat{\mf}^*(\xibf_+)$
and we get
\begin{equation}
\varphi(\xibf_+,\xibf_+) = - \frac 12 \Delta_{\ell\ell}\sigma (\ell , \ell^c) \ebf_\ell \lintprod \tens_2	,
\end{equation}
where the tensor $\tens_2$ is defined as
\begin{equation}
	\tens_2 =  \hat{\mf}(\xibf_+) \odot \hat{\mf}^* (\xibf_+) + \hat{\mf}(\xibf_+) \owedge \hat{\mf}^* (\xibf_+).
	\label{eq:integr-region4}
\end{equation}
As for $\tens_1$, we use the definitions of $\odot$ and $\owedge$ in~\eqref{eq:Tij-int} and~\eqref{eq:Tij-ext} and the identity~\eqref{eq:text-equiv-wedge-int} and we spread out the $ij$-th component of $\tens_2$, which is written
\begin{align}
\tensc_{2,ij} &=
 \Delta_{ii}\Delta_{jj} 
 \bigl( 
\underbrace{
 (\ebf_{i}\lintprod \hat{\mf}(\xibf_+))\cdot (\hat{\mf}^*(\xibf_+)\rintprod \ebf_{j})
			}_{\alpha_{2,ij}} 
  + 
\underbrace{
(-1)^r\Delta_{ij}   \hat{\mf}(\xibf_+) \cdot  \hat{\mf}^*(\xibf_+) 
			}_{\beta_{2,ij}}
	\notag	\\
&\qquad+
\underbrace{
(\ebf_{j}\lintprod\hat{\mf}(\xibf_+))\cdot(\hat{\mf}^*(\xibf_+) \rintprod \ebf_{i}) 
			}_{\alpha_{2,ji}}
\bigr) ,
\label{eq:integr-region5}
\end{align}
and we  substitute the Maxwell field in terms of the potential in the Fourier domain thanks to~\eqref{eq:T_Tij6}, so that we find
\begin{align}
\alpha_{2,ij} &= 4\pi^2
\bigl(\ebf_{i}\lintprod \bigl(\xibf_+ \wedge \hat{\Ab}(\xibf_+)\bigr)\bigr)
\cdot \bigl(\bigl(\xibf_+ \wedge \hat{\Ab}^*(\xibf_+)\bigr)\rintprod \ebf_{j}\bigr),	\label{eq:proof-I2-alpha2} 
\\
\beta_{2,ij} &= 4\pi^2
(-1)^r\Delta_{ij}   \bigl(\xibf_+ \wedge \hat{\Ab}(\xibf_+)\bigr) \cdot  \bigl(\xibf_+ \wedge \hat{\Ab}^*(\xibf_+)\bigr) .
	\label{eq:proof-I2-gamma2}	
\end{align}
We start from~\eqref{eq:proof-I2-alpha2} and we use the relation~\eqref{eq:equiv-lint-wedge} after writing
$\bigl(\xibf_+ \wedge \hat{\Ab}^*(\xibf_+)\bigr)\rintprod \ebf_{j} = (-1)^{r-1} \ebf_{j} \lintprod \bigl(\xibf_+ \wedge \hat{\Ab}^*(\xibf_+)\bigr)$. Thus we get
\begin{multline}
\alpha_{2,ij} = 4\pi^2
(-1)^{r-1}
\bigl((-1)^{r-1} (\ebf_i \cdot \xibf_+) \hat{\Ab}(\xibf_+) + \xibf_+ \wedge \bigl(\ebf_i \lintprod\hat{\Ab}(\xibf_+)\bigr) 
\bigr)
\cdot
\\
\bigl((-1)^{r-1} (\ebf_j \cdot \xibf_+) \hat{\Ab}^*(\xibf_+) + \xibf_+ \wedge \bigl(\ebf_j \lintprod\hat{\Ab}^*(\xibf_+)\bigr) 
\bigr) .
\end{multline}
Again, carrying out all the products, and applying~\eqref{eq:text-equiv-wedge-int} and~\eqref{eq:equiv-lint-wedge}, we obtain
\begin{align}
\alpha_{2,ij} = 4\pi^2
\Bigl(&
(-1)^{r-1} \Delta_{ii}\Delta_{jj}\xi_{+,i}\xi_{+,j}\hat{\Ab}(\xibf_+)\cdot\hat{\Ab}^*(\xibf_+)		\notag
\\
&+\Delta_{ii}\xi_{+,i} \bigl(\ebf_j\lintprod \hat{\Ab}^*(\xibf_+) \bigr)\cdot\bigl(\hat{\Ab}(\xibf_+)\rintprod\xibf_+ \bigr)
\notag	\\
&+ \Delta_{jj}\xi_{+,j} \bigl(\ebf_i\lintprod \hat{\Ab}(\xibf_+) \bigr)\cdot\bigl(\hat{\Ab}^*(\xibf_+)\rintprod\xibf_+ \bigr)
\notag
\\
& -\Bigl(
(-1)^{r-1} (\xibf_+\cdot\xibf_+) \bigl(\ebf_i\lintprod \hat{\Ab}(\xibf_+) \bigr)\cdot \bigl(\ebf_j\lintprod \hat{\Ab}^*(\xibf_+) \bigr)
\notag	\\
&+ \bigl(\xibf_+ \lintprod \bigl(\ebf_i\lintprod \hat{\Ab}(\xibf_+) \bigr)\bigr) \cdot
\bigl(\bigl(\ebf_j\lintprod \hat{\Ab}^*(\xibf_+) \bigr)\rintprod \xibf_+\bigr)
\Bigr)
\Bigr)	.
\end{align}
We note that~\eqref{eq:text-equiv-double-interior} implies
\begin{align}
\bigl(\ebf_j\lintprod \hat{\Ab}^*(\xibf_+) \bigr)\rintprod \xibf_+
=
\ebf_j\lintprod \bigl( \hat{\Ab}^*(\xibf_+) \rintprod \xibf_+ \bigr) = 0		,
\end{align}
and we can simplify $\alpha_{2,ij}$ thanks to the facts that $\xibf_+ \cdot \xibf_+ = 0$ and to the gauge condition $\xibf_+ \lintprod  \hat{\Ab}(\xibf_+)= 0$ from~\eqref{eq:T_Tij8}, obtaining
\begin{equation}
\alpha_{2,ij} = 4\pi^2
(-1)^{r-1} \Delta_{ii}\Delta_{jj}\xi_{+,i}\xi_{+,j}\hat{\Ab}(\xibf_+)\cdot\hat{\Ab}^*(\xibf_+)		.
\end{equation}
Hence, as a consequence,
\begin{equation}
\alpha_{2,ji} = 4\pi^2
(-1)^{r-1} \Delta_{ii}\Delta_{jj}\xi_{+,i}\xi_{+,j}\hat{\Ab}(\xibf_+)\cdot\hat{\Ab}^*(\xibf_+)		.
\end{equation}
Regarding $\beta_{2,ij}$, we write $\xibf_+ \wedge \hat{\Ab}^*(\xibf_+) = (-1)^{r-1}\hat{\Ab}^*(\xibf_+) \wedge  \xibf_+$ and we apply again~\eqref{eq:text-equiv-wedge-int} in~\eqref{eq:proof-I2-gamma2} so that we immediately verify that it vanishes
\begin{align}
\beta_{2,ij} &= - 4\pi^2
\Delta_{ij}
\Bigl(
(-1)^{r-1} (\xibf_+ \cdot \xibf_+)\bigl(\hat{\Ab}(\xibf_+)\cdot\hat{\Ab}^*(\xibf_+)\bigr)
\notag	\\
&\qquad\qquad\qquad\qquad\qquad+ \bigl(\xibf_+ \lintprod\hat{\Ab}(\xibf_+)\bigr)\cdot
\bigl(\hat{\Ab}^*(\xibf_+) \rintprod\xibf_+\bigr)
\Bigr) = 0 .
\end{align}
As a consequence, the result for $\tensc_{2,ij} $ in~\eqref{eq:integr-region5} is
\begin{equation}
\tensc_{2,ij}= 8\pi^2
(-1)^{r-1} \xi_{+,i}\xi_{+,j}\hat{\Ab}(\xibf_+)\cdot\hat{\Ab}^*(\xibf_+)
\end{equation}
and we finally evaluate the tensor $\tens_2$ as
\begin{equation}
\tens_2 = 8\pi^2 \sum_{i\le j} (-1)^{r-1} \xi_{+,i}\xi_{+,j}\bigl\vert \hat{\Ab}(\xibf_+)\bigr\vert^2 \ubf_{ij}	.
\label{eq:integr-region8}
\end{equation}


We move on to the third term of~\eqref{eq:T_Tij3}. The equality $\hat{\mf}(\xibf'_+)=\hat{\mf}^*(\xibf_-)$ allows us to write $\varphi(\xibf_-,\xibf_+')$ as 
\begin{equation}
\varphi(\xibf_-,\xibf_-) = -\frac 12 \Delta_{\ell\ell}\sigma (\ell , \ell^c) \ebf_\ell \lintprod \tens_3	,
\label{eq:integr-region6}
\end{equation}
where we defined the tensor $\tens_3$ as follow
\begin{equation}
	\tens_3 =  \hat{\mf}(\xibf_-) \odot \hat{\mf}^* (\xibf_-) + \hat{\mf}(\xibf_-) \owedge \hat{\mf}^* (\xibf_-).
\label{eq:integr-region7}
\end{equation} 
Comparing~\eqref{eq:integr-region7} and~\eqref{eq:integr-region4}, we note that the difference is only in the presence of $\xibf_-$ instead of $\xibf_+$.
Thus, the mathematical steps are identical, including that the relation $\xibf_+ \cdot\xibf_+ = 0$ has its counterpart $\xibf_- \cdot\xibf_- = 0$.
The gauge conditions in~\eqref{eq:T_Tij8} can also be written as
\begin{equation} \label{eq:gauge2}
\xibf_- \lintprod \hat{\Ab}(\xibf_-)= 0 = \xibf_+ \lintprod \hat{\Ab}^*(\xibf_+).
\end{equation}
So, in analogy with the result obtained in ~\eqref{eq:integr-region8}, the final expression for $\tens_3$ is
\begin{equation}
\tens_3 = 8\pi^2 \sum_{i\le j} (-1)^{r-1} \xi_{-,i}\xi_{-,j}\bigl\vert \hat{\Ab}(\xibf_-)\bigr\vert^2 \ubf_{ij}  .
\label{eq:proofT32}
\end{equation}

We conclude the evaluation of the initial integral in~\eqref{eq:T_Tij3} computing $\varphi(\xibf_-, \xibf_-')$. From ts definition in~\eqref{eq:T_Tij4} and using $\hat{\mf}(\xibf'_-)=\hat{\mf}^*(\xibf_+)$, we get
\begin{equation}
	\varphi(\xibf_-,\xibf_+) = - \frac 12 \Delta_{\ell\ell}\sigma (\ell , \ell^c) e^{-j4\pi\chi_{\ell}x_{\ell} \Delta_{\ell\ell}} \ebf_\ell \lintprod \tens_4 ,
\end{equation}
defined, as in the previous cases,
\begin{equation}
	\tens_4 =  \hat{\mf}(\xibf_-) \odot \hat{\mf}^* (\xibf_+) + \hat{\mf}(\xibf_-) \owedge \hat{\mf}^* (\xibf_+).
\label{eq:integr-region9}
\end{equation}
We can further expand~\eqref{eq:integr-region9} in components which would appear
\begin{align}
\tensc_{4,ij} 
&=
 \Delta_{ii}\Delta_{jj} \bigl( 
\underbrace{ 
 (\ebf_{i}\lintprod \hat{\mf}(\xibf_-))\cdot (\hat{\mf}^*(\xibf_+)\rintprod \ebf_{j}) 
 			}_{\alpha_{4,ij} }
 +
\underbrace{ 
(-1)^r\Delta_{ij}   \hat{\mf}(\xibf_-) \cdot  \hat{\mf}^*(\xibf_+) 
 			}_{\beta_{4,ij} }
 	\notag \\
&\qquad+
\underbrace{
(\ebf_{j}\lintprod\hat{\mf}(\xibf_-))\cdot(\hat{\mf}^*(\xibf_+) \rintprod \ebf_{i})
 			}_{\alpha_{4,ji} }
 \bigr) .
\end{align}
As in the analysis of~$\tensc_{1,ij}$, we express $\alpha_{4,ij}$ and $\beta_{4,ij}$ in terms of the potential
\begin{align}
\alpha_{4,ij} &= 4\pi^2
(-1)^{r-1} (\ebf_{i}\lintprod (\xibf_- \wedge \hat{\Ab}(\xibf_-)))\cdot ( \ebf_{j} \lintprod ( \xibf_+ \wedge\hat{\Ab}^*(\xibf_+))), \label{eq:proof-I4-alpha4} 
\\
\beta_{4,ij} &= 4\pi^2
 (-1)^r\Delta_{ij} (\xibf_- \wedge \hat{\Ab}(\xibf_-)) \cdot  (\xibf_+ \wedge \hat{\Ab}^*(\xibf_+)). \label{eq:proof-I4-beta4} 
\end{align}
We can now note that the differences between~\eqref{eq:integr-region9} and~\eqref{eq:tens1} are in $\xibf_+$ exchanged with $\xibf_-$, $\hat{\mf}(\xibf_+)$ with $\hat{\mf}(\xibf_-)$ and $\hat{\mf}^*(\xibf_-)$ with $\hat{\mf}^*(\xibf_+)$.
So, the differences among~\eqref{eq:proof-I4-alpha4} and~\eqref{eq:proof-I4-beta4} with respect to~\eqref{eq:proof-I1-l1} and~\eqref{eq:proof-I1-l3} are, in addition to the aforementioned, $\hat{\Ab}(\xibf_+)$ interchanged with $\hat{\Ab}(\xibf_-)$ and $\hat{\Ab}^*(\xibf_-)$ with $\hat{\Ab}^*(\xibf_+)$.
We consider~\eqref{eq:T_Tij7} and the gauge condition~\eqref{eq:gauge2}, we replace the conditions~\eqref{eq:xi+Astar-} and~\eqref{eq:xi-A+} with 
\begin{align}
&\xibf_+ \lintprod \hat{\Ab}(\xibf_-) = \xibf_- \lintprod \hat{\Ab}(\xibf_-) + 2\chi_\ell\ebf_\ell \lintprod \hat{\Ab}(\xibf_-)
= 2\chi_\ell\ebf_\ell  \lintprod \hat{\Ab}(\xibf_-) \label{eq:xi+A-}
\\
&\xibf_- \lintprod \hat{\Ab}^*(\xibf_+) = \xibf_+ \lintprod \hat{\Ab}^*(\xibf_+) - 2\chi_\ell\ebf_\ell \lintprod \hat{\Ab}^*(\xibf_+)
= -2\chi_\ell\ebf_\ell \lintprod \hat{\Ab}^*(\xibf_+). \label{eq:xi-Astar+}
\end{align}

If we follow the same procedure applied to obtain $\tens_1$ with the conditions \eqref{eq:gauge2}, \eqref{eq:xi+A-} and~\eqref{eq:xi-Astar+}, 
we can rapidly state 
\begin{equation}
\ebf_\ell \lintprod \tens_4 =0 .
\label{eq:proofT31}
\end{equation}

Then, we write the  integral for the flux in~\eqref{eq:T_Tij3} substituting the definition~\eqref{eq:T_Tij4}. Removing the first and the last summands thanks to \eqref{eq:proofT30} and \eqref{eq:proofT31}, it results
\begin{equation}
\Phi_{\partial \mathcal{V}_\ell^{k+n}} (\Tb)
=
- \frac 18 \Delta_{\ell \ell} \sigma (\ell , \ell^c) \int_{\Delta_{\ell\ell} \xibf_{\bar\ell} \cdot \xibf_{\bar\ell}\le 0}
\drm \xi_{\ell^c} \frac{1}{ \chi_\ell^2}
\ebf_\ell \lintprod 
\bigl(
 \tens_2 + \tens_3
\bigr).
\end{equation}
Using~\eqref{eq:integr-region8} and~\eqref{eq:proofT32}, we get
\begin{multline}
\Phi_{\partial \mathcal{V}_\ell^{k+n}} (\Tb)
=(-1)^{r} \pi^2 
\Delta_{\ell \ell} \sigma (\ell , \ell^c)   \ebf_\ell \lintprod \phantom{.}\\ \int_{\Delta_{\ell\ell} \xibf_{\bar\ell} \cdot \xibf_{\bar\ell}\le 0}
\drm \xi_{\ell^c} \frac{1}{\chi_\ell^2} 
\sum_{i\le j} 
\Bigl(
\xi_{+,i}\xi_{+,j}\bigl\vert \hat{\Ab}(\xibf_+)\bigr\vert^2 
+\xi_{-,i}\xi_{-,j}\bigl\vert \hat{\Ab}(\xibf_-)\bigr\vert^2 
\Bigr)
\ubf_{ij}	.
\label{eq:proofT33}
\end{multline}
We consider that $\chi_\ell^2 = \xi_{+,\ell}^2 = \xi_{-,\ell}^2 $ and we expand the product $\ebf_\ell\lintprod\ubf_{ij}$ thanks to~\eqref{eq:flux_Til} and then use that $\sum_i \xi_{\pm,i}\ebf_i = \xibf_\pm$, so that~\eqref{eq:proofT33} is
\begin{equation} \label{eq:fluxT-v1}
\Phi_{\partial \mathcal{V}_\ell^{k+n}} (\Tb)
= (-1)^{r} \pi^2 \sigma (\ell , \ell^c) \int_{\Delta_{\ell\ell} \xibf_{\bar\ell} \cdot \xibf_{\bar\ell}\le 0}
\drm \xi_{\ell^c} 
\biggl(
\frac{\xibf_{+}}{\xi_{+,\ell}}\bigl\vert \hat{\Ab}(\xibf_+)\bigr\vert^2 
+\frac{\xibf_{-}}{\xi_{-,\ell}}\bigl\vert \hat{\Ab}(\xibf_-)\bigr\vert^2 
\biggr)	.
\end{equation}

As an aside, we may use \cite[p.~184]{Gelfand1966} and that $\chi_\ell = \xi_{+,\ell} = -\xi_{-,\ell}$ to undo the step leading to~\eqref{eq:T_Tij2} to recover the Dirac delta function  
\begin{equation} 
\Phi_{\partial \mathcal{V}_\ell^{k+n}} (\Tb)
= (-1)^{r} 2\pi^2 \sigma (\ell , \ell^c)\int_{-\infty}^{+\infty} \dots \int _{-\infty}^{+\infty}
\drm^{k+n} \xibf \sgn(\xi_\ell)\xibf\bigl\vert \hat{\Ab}(\xibf)\bigr\vert^2\delta(\xibf\cdot\xibf).
\end{equation}

Returning to~\eqref{eq:fluxT-v1}, we split the integral into $I_{\ell,+}+I_{\ell,-}$, where
\begin{equation}
	I_{\ell,\pm} = \int_{\Delta_{\ell\ell} \xibf_{\bar\ell} \cdot \xibf_{\bar\ell}\le 0}
\drm \xi_{\ell^c} \frac{\xibf_{\pm}}{\xi_{\pm,\ell}}\bigl\vert \hat{\Ab}(\xibf_\pm)\bigr\vert^2.
\end{equation}
Taking into account that $\vp(\xb)$ is real, we may express the squared modulus of $\hat{\Ab}(\xibf_-)$ as $\bigl\lvert \hat{\Ab}(\xibf_-)\bigr\rvert^2=\hat{\Ab}(-\xibf_-)\hat{\Ab}^*(-\xibf_-)$. Therefore the integral $I_{\ell,-}$ becomes
\begin{equation}
	I_{\ell,-} = \int_{\Delta_{\ell\ell} \xibf_{\bar\ell} \cdot \xibf_{\bar\ell}\le 0}
\drm \xi_{\ell^c} \frac{\xibf_{-}}{\xi_{-,\ell}} \hat{\Ab}(-\xibf_-)\hat{\Ab}^*(-\xibf_-) .
\end{equation}
Changing the integration variables according to $\xibf_{\bar\ell} \rightarrow \zetabf_{\bar\ell} =-\xibf_{\bar\ell}$, together with the  definition $$\zetabf_\pm = (\zeta_0,\dotsc,\zeta_{\ell-1},\pm \chi_\ell,\zeta_{\ell+1},\dotsc,\zeta_{n+k-1}),$$ yields 
\begin{equation}
I_{\ell,-} 
= 
\int_{\Delta_{\ell\ell} \zetabf_{\bar\ell} \cdot \zetabf_{\bar\ell}\le 0}
\drm \zeta_{\ell^c} 
\frac{\zetabf_{+}}{\zeta_{+,\ell}} \hat{\Ab}(\zetabf_+)\hat{\Ab}^*(\zetabf_+)	.
\label{eq:proofT34}
\end{equation}
Since~\eqref{eq:proofT34} is formally equivalent to $I_{\ell,+}$, the flux can be rewritten as
\begin{equation}
\Phi_{\partial \mathcal{V}_\ell^{k+n}} (\Tb)
=
(-1)^{r} 2\pi^2  \sigma (\ell , \ell^c)
\int_{\Delta_{\ell\ell} \xibf_{\bar\ell} \cdot \xibf_{\bar\ell}\le 0}
\drm \xi_{\ell^c} 
\frac{\xibf_{+}}{\xi_{+,\ell}} \bigl\vert \hat{\Ab}(\xibf_+)\bigr\vert^2	.
\end{equation}

\bibliographystyle{IEEEtran}	
\bibliography{physics.bib}

\end{document}